\documentclass[12pt,letterpaper,english,notitlepage]{revtex4-1}
\usepackage[T1]{fontenc}
\pdfoutput=1
\usepackage[latin9]{inputenc}
\usepackage{amssymb}
\usepackage{graphicx}
\usepackage{esint}

\makeatletter

\usepackage{hyperref}

\makeatother

\usepackage{babel}
\begin{document}
\title{Dynamics of Potential Vorticity Staircase Evolution and Step Mergers
in a Reduced Model of Beta-Plane Turbulence}
\author{M.A. Malkov$^{1}$ and P.H. Diamond$^{1,2}$}
\affiliation{$^{1}$CASS and Department of Physics, University of California, San
Diego~\\
$^{2}$Center for Fusion Sciences, Southwestern Institute of Physics,
Chengdu, Sichuan 610041, People's Republic of China}
\begin{abstract}
A two-field model of potential vorticity (PV) staircase structure
and dynamics relevant to both beta-plane and drift-wave plasma turbulence
is studied numerically and analytically. The model evolves averaged
PV whose flux is both driven by, and regulates, a potential enstrophy
field, $\varepsilon$. The model employs a closure using a mixing
length model. Its link to bistability, vital to staircase generation,
is analysed and verified by integrating the equations numerically.
Long-time staircase evolution consistently manifests a pattern of
meta-stable quasi-periodic configurations, lasting for hundreds of
time units, yet interspersed with abrupt ($\Delta t\ll1$) mergers
of adjacent steps in the staircase. The mergers occur at the staircase
lattice defects where the pattern has not completely relaxed to a
strictly periodic solution that can be obtained analytically. Another
types of stationary solutions are solitons and kinks in the PV gradient
and $\varepsilon$ - profiles. The waiting time between mergers increases
strongly as the number of steps in the staircase decreases. This is
because of an exponential decrease in inter-step coupling strength
with growing spacing. The long-time staircase dynamics is shown numerically
be determined by local interaction with adjacent steps. Mergers reveal
themselves through the explosive growth of the turbulent PV-flux which,
however, abruptly drops to its global constant value once the merger
is completed.
\end{abstract}
\maketitle

\section{Introduction\label{sec:Introduction}}

Pattern and scale selection are omnipresent problems in the dynamics
of fluids and related nonlinear continuum systems. In geophysical
fluids, as described by the beta-plane \citep{Prandtl:2270401} or
quasi-geostrophic equations \citep{pedlosky2013geophysical} - the
mechanisms of formation and scale selection for arrays of jets or
zonal flows \citep{rhines1982homogenization} is of particular interest.
The jets scale constitutes and emergent scale which often defines
the extent of mixing, transport and other important physical phenomena.
Beta-plane and quasi-geostrophic systems evolve by the Lagrangian
conservation of potential vorticity (PV). The latter is an effective
phase space-density which consists of the sum of planetary and fluid
pieces. The question of scale selection then is inexorably wrapped
up in the evolution of \emph{mixing} of potential vorticity. Homogeneous
mixing - predicted by the Prandtl-Batchelor theorem \citep{Prandtl:2270401,pedlosky2013geophysical,rhines1982homogenization},
leads to a uniform PV profile throughout the system with a sharp PV
gradient at the boundary. Inhomogeneous mixing -- linked to bistability
of mixing, multi-scale PV patterns. One of these -- a corrugated
structure called the \emph{potential vorticity staircase }-- is of
particular interest, as it is a long-lived, quasi-stationary pattern
of jets. The struggle between homogenization and (homogeneous mixing)
and corrugation (inhomogeneous mixing) of PV is central to the dynamics
of staircase formation and evolution, which are the foci of this paper. 

The turbulent transport and structure formation phenomenon now commonly
known as a 'staircase', was first understood and described by Philips
\citep{Phillips72}. He considered a density profile in the ocean
that, being stably stratified overall, occasionally reorganizes itself
into layers separated by thin interfaces. The density gradient flattens
in the layers and steepens in the interfaces (sheets). Thus, an initially
linear density profile becomes \emph{ragged}, hence the name 'staircase'.
An interesting aspect of the Phillips paradigm -- by which it can
be distinguished from other pattern formation scenarios in (nonlinear)
unstable media -- is the pre-existing turbulent transport mechanism
that is both supported by, and regulates, the gradient. Even if both
the gradient and flux are initially homogeneous, a small local steepening
(flattening) of the gradient compared with its mean value results
in a local turbulence response that further steepens (flattens) the
gradient. The profile thus undergoes a kind of corrugation instability.
Positive feedback provided by the instability is equivalent to a 'negative
diffusivity' that enhances the profile corrugation instead of smoothing,
in contrast to conventional diffusion. The negative diffusion is often
interpreted as the descending branch on an ``S-curve'' in the flux
- gradient relation, i.e. a range of $\nabla n$ for which $\delta\Gamma/\delta\left(-\nabla n\right)<0$
\citep{Hinton91}. The feedback loop operating macroscopically, drives
the transport supporting turbulence out of the regions with steep
profiles into adjacent regions with flat ones, so as to maintain the
constant net flux across the whole structure, which thus settles at
a \emph{bistable} equilibrium.

Apparently, Philips \citep{Phillips72} did not seek to present his
mechanism as ubiquitous and universally applicable. However, the general
principles behind it suggest looking for its applications to other
media, where similar positive feedback may occur. For example, instead
of negative diffusion, negative viscosity, also resulting from bistability,
may generate strong flow shears. Furthermore, mixing of other quantities,
such as temperature, potential vorticity or salinity may also be considered.
Dritschel and McIntyre \citep{DritMcInt08} review and discuss the
relation between the following three effects: mixing of potential
vorticity, anti-friction effects in horizontal stress, and spontaneous
jet formation \citep{Rhines1994}. Balmforth et al. \citep{Balmforth98}
elaborate on a mathematical model for staircase dynamics by evolving
buoyancy and turbulent energy via two nonlinear diffusion equations,
using a $k-\epsilon$ phenomenology while exploiting an amplitude
and scale dependent mixing length. In a number of respects, our approach
here is along the lines of \citep{Balmforth98} but as the model is
different, so are the results. More about the relation of the present
model to that developed by \citep{Balmforth98} can be found in companion
papers \citep{Ashourvan2017,HahmDiamondTurbTransp2018}.

Apart from fluid mechanics, a promising area for applications of the
staircase concept is plasma transport in magnetic fusion devices,
such as tokamaks. The idea of spontaneously formed transport barriers
(N.B. A staircase may be regarded as a chain of such barriers.) has
attracted significant interest in the fusion community. A transport
bifurcation in the fusion context was first observed in the ASDEX
Tokamak \citep{Wagner82}. This L$\to$H transition occurred with
the formation of a transport barrier at the tokamak edge, following
a (local) transport bifurcation \citep{116_Burrell97,ConnorWilsonLH_REV00,DiamRev05,Garbet08,SchmitzPRL2012,Tynan2013,Gurcan2015}.
Although most of the research on tokamak transport barriers concerned
with such edge phenomena, interest in internal transport barriers
is also significant \citep{Dif-Pradalier2015PhRv,Kosuga2012,Dif-Prad2010}.
Without much risk of oversimplification, a \emph{staircase may be
thought of as a quasiperiodic array of coupled internal transport
barriers}. 

Despite substantial differences in the mechanisms of transport bifurcation
and transport barriers in fluids and plasmas, the apparent commonality
of the phenomena suggests certain general principles behind transport
barrier formation, their subsequent organization into staircases and
the prediction of their possible long-time evolution. We pursue these
goals by using a simple generic staircase model, recently suggested.
In the present paper, we report new results on the following aspects
of the staircase phenomenon. These include
\begin{enumerate}
\item identification of conditions and the parameter space for staircase
formation,
\item the demonstration of staircase persistence by direct numerical integration
of the model equations.
\item finding exact analytic steady state solutions, and exploiting these
for code verification. 
\item the elucidation of staircase dynamics, long time evolution, merger
events and the role of domain boundaries. Special attention is focused
upon the physics of mergers. 
\end{enumerate}
\indent

Taken together, these studies elucidate the basic characteristics
of $\mathbf{E}\times\mathbf{B}$/quasi-geostrophic staircases. 

The plan of the remainder of the paper is as follows. In Sec. \ref{sec:Staircase-model}
we give a summary of the staircase model developed previously for
geostrophic fluids and magnetized plasmas. Sec.\ref{sec:Staircase-Prerequisites}
deals with the bistability conditions, boundary conditions and parameter
regimes required for the staircase formation. Sec.\ref{sec:Staircase-formation}
demonstrates staircase formation by direct numerical integration using
a collocation method. In Sec. \ref{sec:Analytic-solution-for} we
present analytic steady state solutions and use them to verify the
numerical method. In Sec.\ref{sec:Staircase-merger-events} the step
coalescence (merger), long-time dynamics, quasi-equilibrium staircase
configuration and accommodation of boundary conditions are discussed.
Sec.\ref{sec:Discussion-and-conclusions} presents discussion and
conclusions. 

\section{Staircase model\label{sec:Staircase-model}}

The staircase model introduced previously, e.g. \citep{Ashourvan2017},
and applied here to studies of formation and dynamics is relevant
to both geostrophic fluids and magnetized plasmas. These two media
are known to be similar. The present model is one dimensional. This
may appear as a strong limitation, but due to the symmetry of certain
flows (e.g., zonal flows in tokamaks) the 1D approximation is in fact
suitable for the purposes outlined in the Introduction. Note that
useful insights into transport bifurcation observed in tokamaks have
been generated by simple 0D models \citep{KimDiamPRL03,MDhyst09}.
The gain for numerical calculations from such simplifications outweighs
the limitations as it allows much longer integration with sufficient
accuracy. As we will see, staircases typically exhibit disparate spatial
scales, and evolve very rapidly before they reach an asymptotic quasistationary
regime. Moreover, after a long rest such a seemingly steady state
staircase may change abruptly by the merger of two adjacent steps
into one in a matter of a tiny fraction of the rest time. These aspects
of the staircase phenomenon make its studies computationally challenging,
so adaptive mesh refinement is clearly the method of choice here \citep{MuirBacoli2004}.
Three-dimensional numerical studies would be prohibitively expensive
for an accurate determination of the time asymptotic evolution. 

Because of the disparate spatial and time scales discussed above,
code verification is particularly important. A comprehensive code
verification is possible in 1D in time asymptotic regimes by comparison
with exact analytic solutions. Such solutions will be presented below.
Finally, the staircase can be a one-dimensional structure by nature,
so much can be learned about it in 1D. 

The model that we use in this paper is described in detail previously,
so we give only a short review. The model is formulated in terms of
potential vorticity (PV) of a geostrophic fluid \footnote{Discussion of the drift-wave applications of this model in magnetically
confined plasmas is also given in Refs.\citep{Ashourvan2017,HahmDiamondTurbTransp2018}.}, such as the one on the surface of a rapidly rotating planet, i.e.
atmosphere or ocean. This (PV) quantity, $q$, consists of the planetary
vorticity (which we take in the $\beta$-plane approximation) and
fluid vorticity:

\[
q=\beta y+\Delta\psi
\]
where $\psi$ is the stream function, and $y$ is a latitudinal coordinate.
By taking the curl of the Euler equation and adding a forcing term
to its r.h.s. (which we specify later), one can derive the following
equation for $q$

\begin{equation}
\frac{\partial q}{\partial t}-\nabla\psi\times\nabla q=\nu\Delta\psi+f\label{eq:q-generalEq}
\end{equation}
The vector product component perpendicular to the $\beta$-plane is
implied here. Next, we decompose $q$ and $\psi$ into a mean and
fluctuating parts

\[
q=\mbox{\ensuremath{\left\langle q\left(y,t\right)\right\rangle }+\ensuremath{\tilde{q}\left(x,y,t\right)}}
\]
with $\tilde{q}=\Delta\tilde{\psi}$ and substitute this decomposition
into eq.(\ref{eq:q-generalEq}). After separating the $x$-averaged
component $Q\equiv\left\langle q\right\rangle $ from its fluctuating
counterpart squared (enstrophy) $\varepsilon=\left\langle \tilde{q}^{2}\right\rangle /2$,
a familiar closure problem of how to express $\left\langle \nabla\tilde{\psi}\times\nabla\Delta\tilde{\psi}\right\rangle $
through the averaged quantities $\varepsilon$ and $Q$ arises. For
fluctuations which are statistically homogeneous in the $x$-direction
it is straightforward to obtain the following (Taylor) identity for
the $x$-averaged PV flux $\Gamma_{q}$:

\[
-\frac{\partial\Gamma_{q}}{\partial y}\equiv\left\langle \nabla\tilde{\psi}\times\nabla\Delta\tilde{\psi}\right\rangle =\frac{\partial^{2}}{\partial y^{2}}\left\langle \frac{\partial\tilde{\psi}}{\partial x}\frac{\partial\tilde{\psi}}{\partial y}\right\rangle .
\]
The Taylor identity relates the PV flux to the Reynolds stress. Here,
we find it easier to work with PV than with momentum, as PV is locally
conserved. By following the closure prescriptions discussed in previous
papers, we apply a Fickian \emph{Ansatz} for the PV flux:\emph{ }$\Gamma_{q}=-D_{e}\partial Q/\partial y$,
where $D_{e}\left(\varepsilon,Q_{y}\right)$ is the PV diffusivity.
This is assumed to follow a mixing-length hypothesis, $D_{e}\sim l\left|\nabla\tilde{\psi}\right|$,
where $l\left(\varepsilon,Q_{y}\right)$ is the mixing length, introduced
phenomenologically as:

\begin{equation}
\frac{1}{l^{2}}=\frac{1}{l_{0}^{2}}+\frac{1}{l_{R}^{2}}.\label{eq:MixLphen}
\end{equation}
Here, $l_{0}$ is a fixed contribution to the mixing length $l$ that
characterizes the turbulence, e.g., the stirring scale, while $l_{R}$
is the Rhines scale \citep{Rhines1975} at which dissipation of $\varepsilon$
balances its production, so $l_{R}=l_{R}\left(\varepsilon,Q_{y}\right)$.
In turbulent cascades where wave form of energy coexists with turbulent
eddies the Rhines scale is where these two intersect, i.e., where
$k\tilde{v}\sim\omega_{k}$ \citep{Rhines1975}. When the turbulent
energy inverse cascade reaches this scale, it is intercepted and transported
further by waves both in wave-number and configuration space. A macroscopic
consequence of this includes structure formation. A somewhat related
phenomenon is encountered in the context of ``Alfvenization'' of
MHD cascades \citep{goldr97,BeresnLaz2010} in the solar wind and
interstellar medium, where the ``outer scale'' energy is ultimately
converted by waves into thermal and nonthermal plasma energy.

Returning to eq.(\ref{eq:MixLphen}), there is still considerable
freedom in choosing the functional dependence of $l_{R}$ on its arguments.
The only dimensionless combination one may form using the variables
entering eq.(\ref{eq:MixLphen}) is $l_{0}^{2}Q_{y}^{2}/\varepsilon\equiv l_{0}^{2}/l_{R}^{2}$.
So, we may slightly generalize the relation in eq.(\ref{eq:MixLphen})
and write $l_{0}/l=\left(1+l_{0}^{2}Q_{y}^{2}/\varepsilon\right)^{\kappa}$.
We choose $\kappa=2$ and will comment on this choice in Sec.\ref{sec:Staircase-Prerequisites}.
Replacing the eddy velocity in the Fick's law by $l_{0}\sqrt{\varepsilon}$
and measuring $y$ in units of $l_{0},$ we can write the averaged
Eq.(\ref{eq:q-generalEq}) as follows 

\begin{equation}
Q_{t}=\partial_{y}\frac{\varepsilon^{1/2}}{\left(1+Q_{y}^{2}/\varepsilon\right)^{2}}Q_{y}+DQ_{yy}\label{eq:QeqDim}
\end{equation}
Here we added to the eddy diffusivity (the first term on the rhs),
a conventional collisional diffusivity $D$ that may be associated
with the molecular viscosity $\nu$ in eq.(\ref{eq:q-generalEq}).
We prefer to consider it as a modest additive regularization of the
turbulent diffusivity, $D$. Applying similar argument to the turbulent
part of PV and adding the terms responsible for its production, damping
and unstable growth (see \citep{Ashourvan2017,HahmDiamondTurbTransp2018}
for further details), we write an evolution equation for the potential
enstrophy $\varepsilon$ as follows: 

\begin{equation}
\varepsilon_{t}=\partial_{y}\frac{\varepsilon^{1/2}}{\left(1+Q_{y}^{2}/\varepsilon\right)^{2}}\varepsilon_{y}+D\varepsilon_{yy}+\frac{\varepsilon^{1/2}}{\left(1+Q_{y}^{2}/\varepsilon\right)^{2}}Q_{y}^{2}-\frac{\varepsilon^{3/2}}{\varepsilon_{0}}+\gamma\sqrt{\varepsilon}\label{eq:EpsEqDim}
\end{equation}
For the purposes of regularization, we use the same background diffusivity
$D$ as in eq.(\ref{eq:QeqDim}) (see also below), $\gamma$ is the
strength of the forcing, while $\varepsilon_{0}$ quantifies the nonlinear
damping of the enstrophy. Apart from these three parameters, the problem
depends on the domain size in $y$- direction. Let us measure it in
the units of $l_{0}$ and denote by $L.$ Altogether, the system thus
depends on four parameters ($D,\varepsilon_{0},\gamma$ and $L$),
of which one can be removed by re-scaling the variables. This reduction
is, in fact, crucial to the search for staircase solutions, as they
occupy a small domain in parameter space. So, by replacing

\begin{equation}
\varepsilon\to\gamma\varepsilon,\;\;\;Q\to\sqrt{\gamma}LQ,\;\;\;y\to Ly,\;\;\;t\to\gamma^{-1/2}L^{2}t\label{eq:Units}
\end{equation}
eqs.(\ref{eq:QeqDim}) and (\ref{eq:EpsEqDim}) transform to the following
system

\begin{equation}
Q_{t}=\partial_{y}\frac{\varepsilon^{1/2}}{\left(1+Q_{y}^{2}/\varepsilon\right)^{2}}Q_{y}+DQ_{yy}\label{eq:QeqNdim}
\end{equation}

\begin{equation}
\varepsilon_{t}=\partial_{y}\frac{\varepsilon^{1/2}}{\left(1+Q_{y}^{2}/\varepsilon\right)^{2}}\varepsilon_{y}+D\varepsilon_{yy}+L^{2}\left\{ \frac{Q_{y}^{2}}{\left(1+Q_{y}^{2}/\varepsilon\right)^{2}}-\frac{\varepsilon}{\varepsilon_{0}}+1\right\} \varepsilon^{1/2}\label{eq:EpsEqNdim}
\end{equation}
and the integration domain is now $y\in\left[0,1\right]$. These are
strongly nonlinear driven/damped parabolic equations possessing numerous
stationary and time-dependent solutions. In the next section, we discuss
the strategy of our search for the solutions with required staircase
properties. To conclude this section, we consider the total enstrophy
budget by introducing this quantity as

\[
\mathcal{E}\equiv\intop_{0}^{1}\left(Q^{2}/2+\varepsilon/L^{2}\right)dy.
\]
From eqs.(\ref{eq:QeqNdim}) and (\ref{eq:EpsEqNdim}), we obtain

\[
\frac{d}{dt}\mathcal{E}=\intop_{0}^{1}\sqrt{\varepsilon}\left(1-\frac{\varepsilon}{\varepsilon_{0}}\right)dy+\left.\frac{\sqrt{\varepsilon}\left(\varepsilon_{y}/L^{2}+QQ_{y}\right)}{\left(1+Q_{y}^{2}/\varepsilon\right)^{2}}\right|_{0}^{1}-D\left(\intop_{0}^{1}Q_{y}^{2}dy+\left.QQ_{y}\right|_{0}^{1}\right)
\]
It is seen that the volumetric enstrophy production due to the instability
(first term under the first integral) can be balanced by the nonlinear
damping (second term under the integral). The possible enstrophy leak
through the boundaries (second term) and small diffusive dissipation
can also be compensated by adjusting the nonlinear dissipation rate,
$\propto\varepsilon_{0}^{-1}$. 

\section{Staircase Prerequisites \label{sec:Staircase-Prerequisites}}

A numerical integration of eqs.(\ref{eq:QeqNdim}-\ref{eq:EpsEqNdim}),
if not thoroughly planned, shows that staircases (SC) are not ubiquitous
solutions that arise from almost any randomly chosen set of parameters
and initial conditions. On the contrary, one needs to search for them
carefully in a multidimensional parameter space. However, as we will
see from the sequel, once the appropriate corner in the parameter
space is located, the SC solutions arise as remarkably robust asymptotic
attractors of the system given by eqs.(\ref{eq:QeqNdim}-\ref{eq:EpsEqNdim}).
Apart from the three parameters directly entering these equations
($D,L,\varepsilon_{0}$), two or three additional parameters enter
from the boundary conditions for $Q$ and $\varepsilon$, depending
on assumptions, discussed briefly below. 

First, throughout this paper we impose Dirichlet boundary conditions
but fix a PV contrast across the domain, thus maintaining a constant
average flux. Specifically, we assume a SC structure to form in a
limited $y$-domain (in our variables $y\in\left[0,1\right]$). On
each side of this domain a stationary level of $\varepsilon$ is maintained,
which in most of the cases considered in this paper is the same: $\varepsilon\left(0,1\right)=\varepsilon_{{\rm B}}$.
This reduces the total number of parameters by one. Next, the PV,
$Q$, is determined up to an arbitrary constant, so we fix its value
on one end, $Q\left(0\right)=0$. Then we set $Q\left(1\right)=Q_{{\rm B}}$,
so the enstrophy inside the SC is driven by the gradient $Q_{y}$
(the first term in braces in eq.{[}\ref{eq:EpsEqNdim}{]}, with $\left\langle Q_{y}\right\rangle =Q_{{\rm B}}$),
in addition to the constant drive given by the very last term in the
braces. At a minimum, we thus have a 5D parameters space to search
in for a SC solution. Clearly, we need directions for our search. 

In rough terms, a SC structure described in Sec.\ref{sec:Introduction}
may result from the loss of stability of a ground state solution of
eqs.(\ref{eq:QeqNdim}-\ref{eq:EpsEqNdim}) characterized by the constant
values $\varepsilon=\varepsilon_{B}$ and $Q_{y}=Q_{B}$ that annihilate
the term in braces in eq.(\ref{eq:EpsEqNdim}). Then, nonlinear saturation
of the instability may possibly lead to a (quasi)stationary SC solution.
A generic paradigm here is \emph{bistability}, \emph{where apart from
the unstable ground state, there exist two stable steady states}.
The system jumps to one of these when the ground state becomes unstable.
To explain the conditions for this scenario, let us denote the enstrophy
production-dissipation term in braces on the r.h.s. of eq.(\ref{eq:EpsEqNdim})
as 

\begin{equation}
R\equiv\frac{Q_{y}^{2}}{\left(1+Q_{y}^{2}/\varepsilon\right)^{2}}-\frac{\varepsilon}{\varepsilon_{0}}+1\label{eq:RofEps}
\end{equation}

A SC, particularly the one with a large number of steps, evidently
requires $L\gg1$. Otherwise, the diffusive terms in eqs.(\ref{eq:QeqNdim}-\ref{eq:EpsEqNdim})
will dominate, thus driving the $\varepsilon$ and $Q_{y}$ profiles
to constants. Therefore, assuming $\partial_{t}Q\approx0$ and $L\to\infty$,
it follows that $R\to0$. Thus, with some reservations discussed below,
eq.(\ref{eq:EpsEqNdim}) can be written for $L\gg1$ as

\begin{equation}
\varepsilon_{t}=L^{2}\sqrt{\varepsilon}R\left(\varepsilon,Q_{y}\right)\label{eq:EpsbytTrunc}
\end{equation}
It should be noted here that (as $L\gg1$) the first two terms on
the r.h.s. of eq.(\ref{eq:EpsEqNdim}) are small terms, with higher
derivative discarded in eq.(\ref{eq:EpsbytTrunc}). Boundary layers
are expected between the states with high and low values of $\varepsilon$
associated with two stable fixed points in eq.(\ref{eq:EpsbytTrunc}).
Also note that $Q_{y}$ may (and will) jump between these fixed points
along with $\varepsilon$, but the jump description requires treating
the neglected higher derivative terms that will be taken into account
in Sec.\ref{sec:Analytic-solution-for}. We will call the thin regions
over which $\varepsilon$ and $Q_{y}$ jump, the 'corners', as they
appear as such in the $Q\left(y\right)$ profile (see Fig.\ref{fig:1StepSC}).
A region of flat $Q$ (large $\varepsilon$) attached to the corner
on one side we call a 'step'. A contrasting region, where $Q$ is
steep (small $\varepsilon),$ we call a 'shear layer' or 'jump'. In
essence, such structure corresponds well to the SC phenomenon, since
the corners between the shear layers and steps, as envisioned by Phillips
\citep{Phillips72}, are simply the internal boundary layers.
\begin{figure}
\includegraphics[bb=0bp 100bp 612bp 700bp,scale=0.5,angle=270]{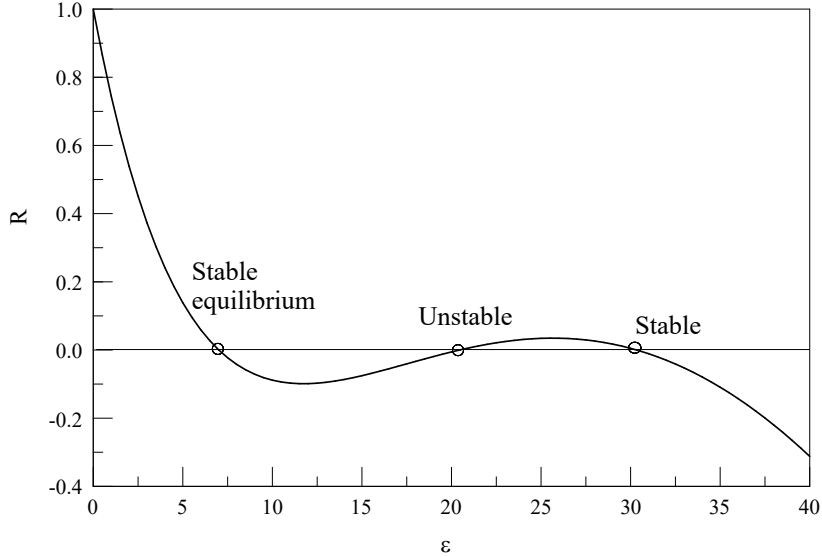}

\caption{Enstrophy production-dissipation term $R$ in eqs.(\ref{eq:EpsEqNdim},\ref{eq:RofEps})
as a function of enstrophy $\varepsilon$, shown for a fixed mean
vorticity gradient $Q_{y}=35$ and $\varepsilon_{0}=3.55$. \label{fig:Enstrophy-production-dissipation}}
\end{figure}

\begin{figure}
\includegraphics[bb=-50bp 100bp 550bp 750bp,scale=0.5,angle=270]{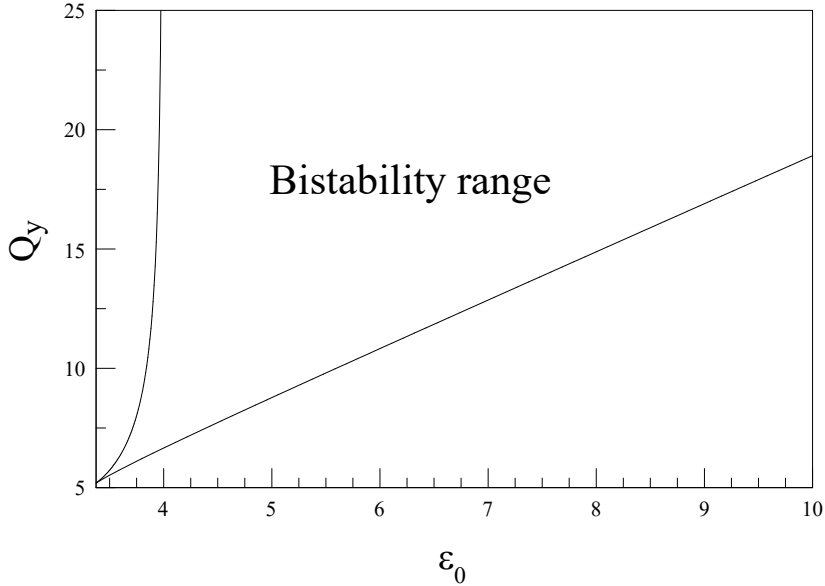}

\caption{Part of parameter space in variables $\varepsilon_{0},Q_{y}$ where
the SC solutions are possible.\label{fig:SCparamSpace}}
\end{figure}

The conditions for generating the SC structures described above can
be approached assuming constant local enstrophy and vorticity gradient
$\varepsilon,$$Q_{y}\equiv const$. Here, we constrain the parameters
in the driving term $R$ to have a bistable form shown in Fig.\ref{fig:Enstrophy-production-dissipation}.
In this case, the alternating layers correspond to jumps between two
stable equilibria (fixed points), over an unstable one. Of course,
one can directly solve the cubic relation $R=0$ for, e.g., $\varepsilon$
as a function of $Q_{y}$ and $\varepsilon_{0}$, thus determining
conditions for the three roots to exist. However, this approach is
algebraically tedious, so we take a different route. First, denote
$\xi\equiv\varepsilon/Q_{y}^{2}=l_{R}^{2}$ and rewrite the equation
$R=0$ after dividing it by $Q_{y}^{2}$ 
\begin{figure}
\includegraphics[bb=0bp 200bp 612bp 700bp,clip,scale=0.55]{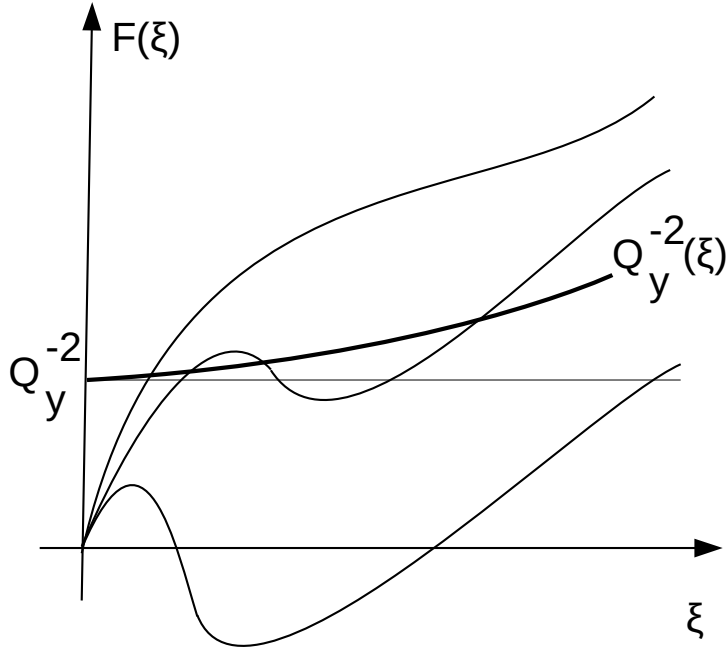}

\caption{Left-hand side of eq.(\ref{eq:ExtremaFofKsi}), $F\left(\xi\right),$
schematically drawn for three different values of parameter $\varepsilon_{0}$.\label{fig:LHSofFOfKsi}}
\end{figure}
\begin{equation}
F\left(\xi\right)\equiv\frac{\xi}{\varepsilon_{0}}-\frac{\xi^{2}}{\left(1+\xi\right)^{2}}=\frac{1}{Q_{y}^{2}}\label{eq:ExtremaFofKsi}
\end{equation}
In this form, the variables $\xi$ and $Q_{y}$ are separated while
$\varepsilon_{0}$ is considered as a constant parameter. For this
equation to have three real roots (the scenario of bistability), the
value $1/Q_{y}^{2}$ must fall between the local extrema of $F\left(\xi\right)$,
provided that they exist, i.e.:
\begin{figure}
\includegraphics[bb=100bp 150bp 612bp 792bp,scale=0.6]{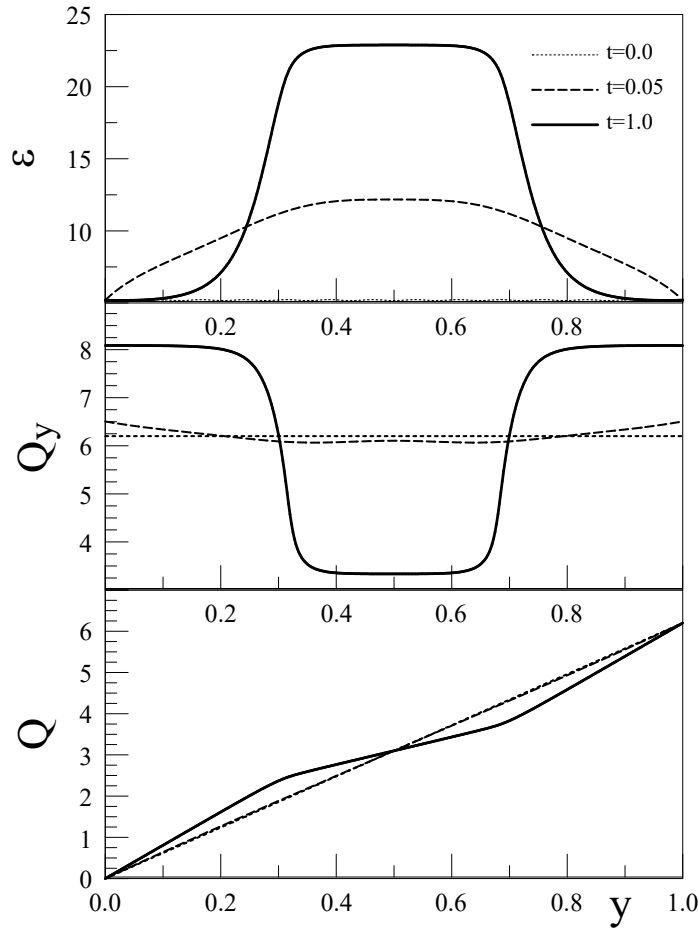}

\caption{Generation of a one-step profile out of an unstable ground state superposed
by small perturbations (short-dash lines). Perturbations are low-amplitude,
so short scales are not seen clearly in the initial profile but they
do not affect the profile evolution significantly. The Dirichlet boundary
conditions are applied for $Q$ and $\varepsilon$ at both boundaries.
The other parameters are $\varepsilon_{0}$=3.8, $L^{2}=2.3\cdot10^{3}$,
$D=1.5$, the boundary conditions can be read from the plots.\label{fig:1StepSC}}
\end{figure}
 
\[
F_{{\rm min}}<Q_{y}^{-2}<F_{{\rm max}}.
\]
(see Fig.\ref{fig:LHSofFOfKsi}). Denoting the points of extrema by
$\xi_{0,1}$, this constraint can be written, using eq.(\ref{eq:ExtremaFofKsi})
as follows

\begin{equation}
\xi_{1}\left(1-\xi_{1}\right)<\frac{2\varepsilon_{0}}{Q_{y}^{2}}<\xi_{0}\left(1-\xi_{0}\right)\label{eq:QyBistInt1}
\end{equation}
Note that the left term of this inequality becomes negative for sufficiently
large $\varepsilon_{0}$ (lower curve in Fig.\ref{fig:LHSofFOfKsi}).
The two extremal points of $F\left(\xi\right),$ $\xi_{0}<\xi_{1}$
, where $F^{\prime}\left(\xi_{0,1}\right)=0,$ are also to be found
from a cubic equation, but this equation is much simpler than the
original one, given by eq.(\ref{eq:ExtremaFofKsi}). The requirement
for the two isolated roots $\xi_{0,1}$ to exist is $\varepsilon_{0}$>27/8
(Appendix \ref{sec:Appendix1}). This condition proved very useful
in the search for a SC regime. However, the latter is not precise
in that $Q_{y}$ and $\xi$ also change during the transition from
one stable state to the other. In fact, one can easily determine $Q_{y}\left(\xi\right)$
variation by assuming that it changes in space but remains stationary.
This assumption relates $Q_{y}$ to $\xi$ by the constant diffusive
flux in eq.(\ref{eq:QeqNdim}). Denoting the flux of $Q$ by $b$
(see Appendix \ref{sec:Appendix1} and eq.{[}\ref{eq:Qfluxb}{]} below)
we obtain the following relation for $Q_{y}$ to be used in the constraint
(\ref{eq:QyBistInt1}) for connecting the two stable roots of eq.(\ref{eq:ExtremaFofKsi})

\begin{equation}
\frac{1}{Q_{y}}=\frac{D}{2b}+\sqrt{\frac{D^{2}}{4b^{2}}+\frac{\xi^{5/2}}{b\left(1+\xi\right)^{2}}}\label{eq:OneOverQy}
\end{equation}
By substituting the last relation into the rhs of eq.(\ref{eq:ExtremaFofKsi}),
one can solve it for $\xi$ in terms of the three parameters $\varepsilon_{0},D,$
and $b$. For certain values of these parameters, three isolated roots
are possible, of which the largest and the smallest correspond to
neighboring layers in a staircase solution, or to the two stable roots
of the truncated equation (\ref{eq:EpsbytTrunc}). The intermediate
root is unstable. For practical reasons, instead of locating all three
roots, we constrain the parameter space by the simple analytic formulae
(\ref{eq:QyBistInt1}) and ( \ref{eq:KsiN}). They provide a range
of $Q_{y}$ for possible staircase solutions in terms of $\varepsilon_{0}$.
The region in the $\varepsilon_{0},Q_{y}$ plane, where to look for
the staircase numerically, is shown in Fig.\ref{fig:SCparamSpace}.
It follows that a stationary staircase structure is a quasi-periodic
sequence of regions with alternating upper and lower stable $\varepsilon$
values in Fig.\ref{fig:Enstrophy-production-dissipation}. As we will
see in the sequel, time-asymptotically this solution can be calculated
analytically. The exact analytic solution provides both a guidance
in exploring the time-dependent regimes and an excellent code verification
tool. 

In addition to the above guidance, the following consideration has
proven useful in search for staircase solutions. As they result from
an unstable stationary solution with initially constant $\varepsilon$
and $Q_{y}$, a stability analysis of the full system can be performed.
This replaces the local analysis above, based on the zeroes of the
function $R\left(\varepsilon\right)$ and signs of its derivatives
$R^{\prime}\left(\varepsilon\right)$. This extended analysis, even
if it probes only simple perturbations of the type $\propto\exp\left(iky-i\omega t\right)$,
is rather tedious. It is also more restrictive, as it does not capture
the bistable state described earlier. So, we do not reproduce the
linear analysis here. However, a broader insight into the parameter
choice for obtaining the staircase solution has been gained from this
analysis. As mentioned earlier, the mixing length scaling in eq.(\ref{eq:MixLphen})
can be written more generally as $l_{0}^{2}/l^{2}=\left(1+l_{0}^{2}/l_{R}^{2}\right)^{\kappa}$
and the stability analysis can be performed for this, more general
form of the mixing length. Our particular choice $\kappa=2$ in the
model equations (\ref{eq:QeqNdim}-\ref{eq:EpsEqNdim}) was precisely
dictated by the instability condition for a steady state solution
with constant $\varepsilon$ and $Q_{y}$. At the same time, even
without performing such a stability analysis of the full system, an
alternative choice $\kappa=1$ instead of $\kappa=2$ may be shown
to be inconsistent with SC solutions. Namely, the denominator of the
first term in $R$ from eq.(\ref{eq:RofEps}) is simply $1+Q_{y}^{2}/\varepsilon$
in this case. Therefore, the function $R\left(\varepsilon\right)$
has only one positive stable root at which $R^{\prime}\left(\varepsilon\right)<0$.
Hence, no bistability occurs, so no staircase forms!

Another important aspect of the stability analysis regards the possible
number of steps in the structure. Indeed, by contrast to the above
discussed \emph{local} bistability based on eq.(\ref{eq:EpsbytTrunc})
(in essence, a $k\to0$ limit), the standard linear analysis assumes
perturbations of the form $\propto\exp\left(-i\omega t+iky\right)$.
The small scales are damped, as $\Im\omega\simeq-k^{2}$, but the
growth rate turns positive for small $k$ and sufficiently large $L$.
Thus, there must be a maximum unstable $k$, and possibly even maximum
$\Im\omega\left(k\right)$, depending on the boundary conditions.
\emph{This} $k$ \emph{sets the number of steps in the staircase!}
However, as numerical integration shows, \emph{this initial number
}(being also somewhat sensitive to the initial conditions) \emph{quickly
relaxes to a smaller number of steps, as the staircase grows to a
nonlinear level}. This quasi-equilibrium staircase configuration and
its time evolution are the main focus of our study below. 

\section{Staircase formation\label{sec:Staircase-formation}}

It follows from the above considerations that the number of steps
in a staircase must grow with the parameter $L$. It is thus natural
to assume that at some moderate value of $L$, this number can be
as small as one. Shown in Fig.\ref{fig:1StepSC} is a one-step profile
generated from an unstable ground state with small scale weak perturbations
(short-dash lines) superposed. The initially small scales are quickly
damped, and the system evolves to a one-step profile, shown by heavy
lines. In Sec.\ref{sec:Analytic-solution-for} we will demonstrate
that this profile coincides with an exact stationary solution of the
system. As it appears to be stable, it must persist indefinitely,
thus constituting a single step attractor. The $\varepsilon$ and
$Q_{y}$ profiles are symmetric with respect to the mid-plane, as
the boundary conditions also are, $\varepsilon\left(0\right)=\varepsilon\left(1\right)$
.

\begin{figure}
\includegraphics[bb=100bp 150bp 612bp 792bp,clip,scale=0.6]{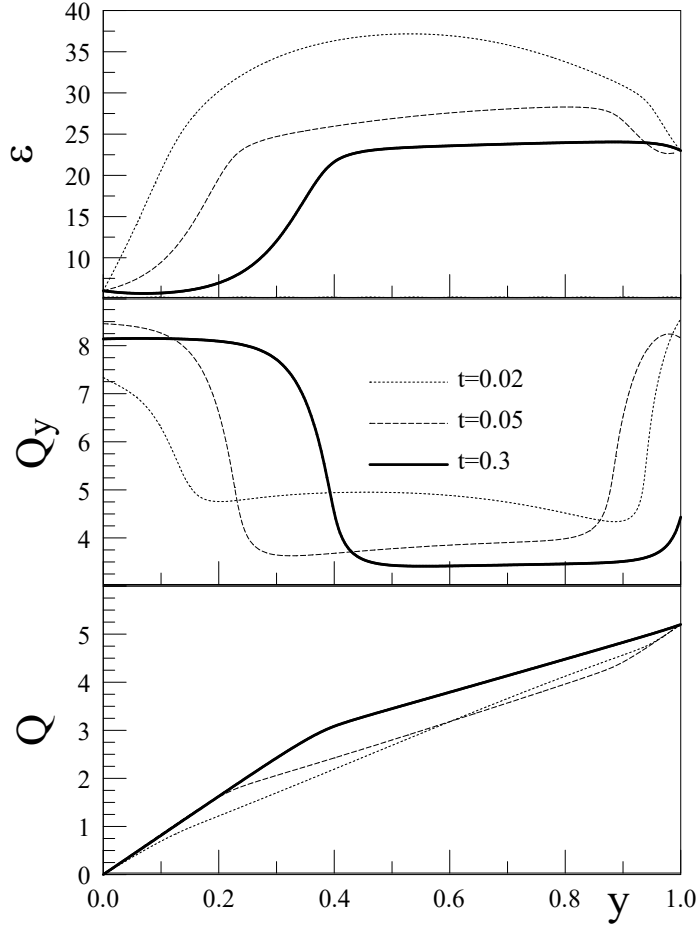}

\caption{The same plots as in Fig.\ref{fig:1StepSC} but for $L^{2}=900$.
Other parameters are the same, except the turbulence level is kept
at different levels on right and left boundary. As the mid-plane symmetry
is broken, the initially formed profiles propagate to the right boundary.
After $t\lesssim1$, the $Q$-profile relaxes to one step at the right
boundary and a shear layer at the left. This configuration persists
in time, as the boundary conditions are consistent with a part of
periodic stationary solution, Sec.\ref{sec:Analytic-solution-for},
that fits into the integration domain.\label{fig:MovingStep}}
\end{figure}

An example of asymmetric profile is shown in Fig.\ref{fig:MovingStep}.
In this case, the boundary conditions are different at the left and
right boundary, $\varepsilon\left(0\right)\neq\varepsilon\left(1\right)$.
Although the system also evolves into a staircase with only one step,
this time the step attaches itself to one of the boundaries. The steep
part of the $Q$-profile attaches itself to the opposite boundary.
A point of special interest associated with this simple configuration
is its similarity to the temperature and density distributions in
H-mode (high confinement regime) of operation of magnetic fusion confinement
devices. The region at the left boundary with a steep gradient of
the transported (by turbulence) quantity ($Q$ in this case) is characterized
by transport reduction associated with the low turbulence level, ($\varepsilon$
here).

Now that we have demonstrated that a stationary staircase is indeed
a strong attractor for the time-dependent system given by eqs.(\ref{eq:QeqNdim}-\ref{eq:EpsEqNdim}),
it is worthwhile to investigate all possible steady-state solutions
analytically. This investigation will be useful in numerical studies
of staircase dynamics, presented later in Sec.\ref{sec:Staircase-merger-events}.

\section{Analytic solution for time-asymptotic staircase\label{sec:Analytic-solution-for}}

Analytical time-asymptotic staircase solutions can be obtained easily.
These solutions are also relevant to the staircase dynamics since
-- as we will see from the numerical studies -- a typical multi-step
staircase does not change in time, apart from quick merger events.
One may refer to it as ``meta-stationary''. Between such merger
events, the staircase is perfectly described by analytic solutions
that we obtain below and compare against numerical solutions in Sec.\ref{subsec:Comparisonasymptotic}. 

Assuming a steady state, from eq.(\ref{eq:QeqNdim}) we deduce

\begin{equation}
\left[\frac{\varepsilon^{1/2}}{\left(1+Q_{y}^{2}/\varepsilon\right)^{2}}+D\right]Q_{y}\equiv b=const\label{eq:Qfluxb}
\end{equation}
Instead of $Q$ and $y$, it is convenient to use the following two
variables as dependent and independent, respectively 

\begin{equation}
\psi=\frac{Q_{y}}{\sqrt{\varepsilon}},\;\;\;\eta=\sqrt{2}Ly,\label{eq:PsiEtaDef}
\end{equation}
We keep $\varepsilon$ as the second dependent variable. Assuming
also $\partial\varepsilon/\partial t=0$, multiplying eq.(\ref{eq:EpsEqNdim})
by a factor $2\varepsilon_{y}/bQ_{y}$ and integrating once in $y$,
we arrive at the following first integral of this equation

\begin{equation}
\frac{\varepsilon_{\psi}^{2}}{\psi^{2}\varepsilon}\left(\frac{d\psi}{d\eta}\right)^{2}+W\left(\psi\right)=E=const\label{eq:FirstInt}
\end{equation}
where

\begin{eqnarray}
W\left(\psi\right) & = & \varepsilon-\frac{2D}{3b}\psi\varepsilon^{3/2}+\frac{\varepsilon}{b\psi}\left(1-\frac{\varepsilon}{2\varepsilon_{0}}\right)\nonumber \\
 & + & \frac{2D}{3b}\int\varepsilon d\psi+\frac{1}{b}\int\frac{\varepsilon}{\psi^{2}}\left(1-\frac{\varepsilon}{2\varepsilon_{0}}\right)d\psi,\label{eq:WofPso}
\end{eqnarray}
$\varepsilon_{\psi}=\partial\varepsilon/\partial\psi$ and $\varepsilon\left(\psi\right)$
can now be written in the following explicit form

\begin{equation}
\varepsilon=\left(1+\psi^{2}\right)^{2}\left[\sqrt{\frac{b}{\psi}+\frac{1}{4}D^{2}\left(1+\psi^{2}\right)^{2}}-\frac{1}{2}D\left(1+\psi^{2}\right)^{2}\right]^{2}\label{eq:EpsOfPsi}
\end{equation}
It follows that the first integral in eq.(\ref{eq:FirstInt}) provides
the steady state solution of eq.(\ref{eq:EpsEqNdim}) in the form
of $\psi\left(\eta\right)$ which can be obtained by inverting the
function $\eta\left(\psi\right)$. This in turn, can be derived from
eq.(\ref{eq:FirstInt}) by quadrature. Furthermore, using eqs.(\ref{eq:PsiEtaDef})
and (\ref{eq:EpsOfPsi}), one can obtain the steady state solution
in original variables, $Q\left(y\right)$ and $\varepsilon\left(y\right)$.

For the purpose of comparison of the solution given by eq.(\ref{eq:FirstInt})
with the asymptotic regime obtained from the numerical integration
of eqs.(\ref{eq:QeqNdim}-\ref{eq:EpsEqNdim}), we return to the original
coordinate $y$ and write the solution in the form of $y=y\left(\psi\right):$ 

\begin{equation}
y=\frac{1}{\sqrt{2}L}\int\frac{\partial\varepsilon}{\partial\psi}\frac{d\psi}{\psi\sqrt{\varepsilon\left[E-W\left(\psi\right)\right]}}\label{eq:Ysol}
\end{equation}
Apart from an arbitrary constant $y_{0}$, that can always be added
to the rhs to adjust the position of the staircase in $y$, the solution
$\psi\left(y\right)$ in eq.(\ref{eq:Ysol}) depends on two further
constants, $E$ and $b$. The latter is the flux of $Q$ that enters
$W$ in the last expression by virtue of eq.(\ref{eq:WofPso}). Although
$b$ is related to the boundary conditions because of eq.(\ref{eq:Qfluxb}),
under the Dirichlet boundary conditions employed here $Q_{y}$ is
not fixed at the boundaries. Therefore, $b$ becomes constant only
when the system reaches a steady (or meta-stationary, as discussed
earlier) state. Before such state, presented in eq.(\ref{eq:Ysol}),
is reached $b$ changes in time. 
\begin{figure}
\includegraphics[bb=100bp 200bp 612bp 792bp,clip,scale=0.55]{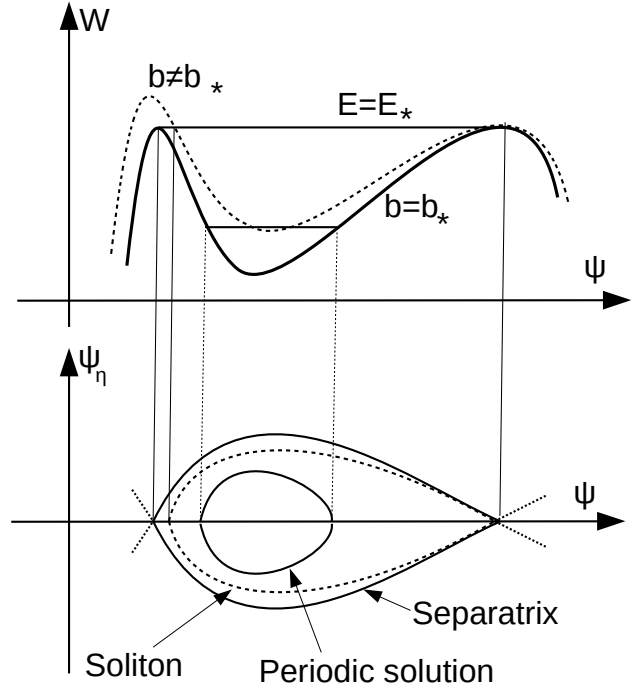}

\caption{'Oscillator's' pseudo-potential and its phase plane. The top panel
shows the function $W\left(\psi,b\right)$ from eq.(\ref{eq:FirstInt})
for the case when two maxima that correspond to two zeros of function
$R\left(\varepsilon\right)$ in eq.(\ref{eq:RofEps}) are at the same
value of $W=E_{{\rm max}}.$ The maxima are the same for a specific
value of the constant flux $b$ in eq.\ref{eq:QeqNdim}. For $E<E_{\max}$
solutions are periodic.\label{fig:'Oscillator'-pseudo-potential-an}}
\end{figure}
The role of the constants $E$ and $b$ can be elucidated by returning
to the first integral in eq.(\ref{eq:FirstInt}) and interpreting
it as a constant total energy of a pendulum with a variable mass (coefficient
in front of $\psi_{\eta}^{2}$) moving in a potential well $W\left(\psi\right)$.
The variable $\psi$ plays the role of a coordinate here while $\eta$
is 'time'. One form of $W\left(\psi\right)$, shown in Fig.\ref{fig:'Oscillator'-pseudo-potential-an},
corresponds to a specific value of $b=b_{*}$ for which the two maxima
of $W\left(\psi\right)$ are equal. This is an important case since
it admits heteroclinic orbits connecting two hyperbolic points of
the 'pendulum' at a specific value of $E=E_{*}$. The orbits correspond
to the two branches of a separatrix shown on the phase plane in Fig.\ref{fig:'Oscillator'-pseudo-potential-an}.
These particular values of $b=b_{*}$ and $E=E_{*}$ correspond to
an isolated transition from low to high values of $\psi$, when $y$
runs from $-\infty$ to $+\infty$. The original variables $\varepsilon$
and $Q_{y}$ can always be restored unambiguously from $\psi\left(y\right)$
using eqs.(\ref{eq:EpsOfPsi}) and (\ref{eq:PsiEtaDef}). The mirror
branch of this orbit corresponds to the reverse transition and fixed
points correspond to the two stable roots of the function $R\left(\varepsilon\right)$
introduced in Sec.\ref{sec:Staircase-Prerequisites} and depicted
in Fig.\ref{fig:Enstrophy-production-dissipation}. Clearly, for a
heteroclinic orbit to exist the two areas cut by the abscissa from
$R\left(\varepsilon\right)$ curve between the stable and two unstable
roots must satisfy a certain relation. This is that measured (integrated)
in the variable $y$ instead of $\varepsilon$, these areas must be
equal, as it follows from the derivation of eq.(\ref{eq:FirstInt}).
Using the pendulum analogy again, the heteroclinic orbit connects
two unstable equilibria (the two humps on the potential energy profile).
Therefore, the accelerating phase of the trajectory must be exactly
annihilated by the decelerating phase. This is equivalent to the familiar
Maxwell's construction, illustrated in Fig.\ref{fig:Maxwell}. The
Maxwell rule is common for other transition phenomena, both flux and
source driven \citep{LebDiam97,murray2001mathematical,MDtranspBif08}.
\begin{figure}
\includegraphics[bb=0bp 400bp 612bp 792bp,clip,scale=0.65]{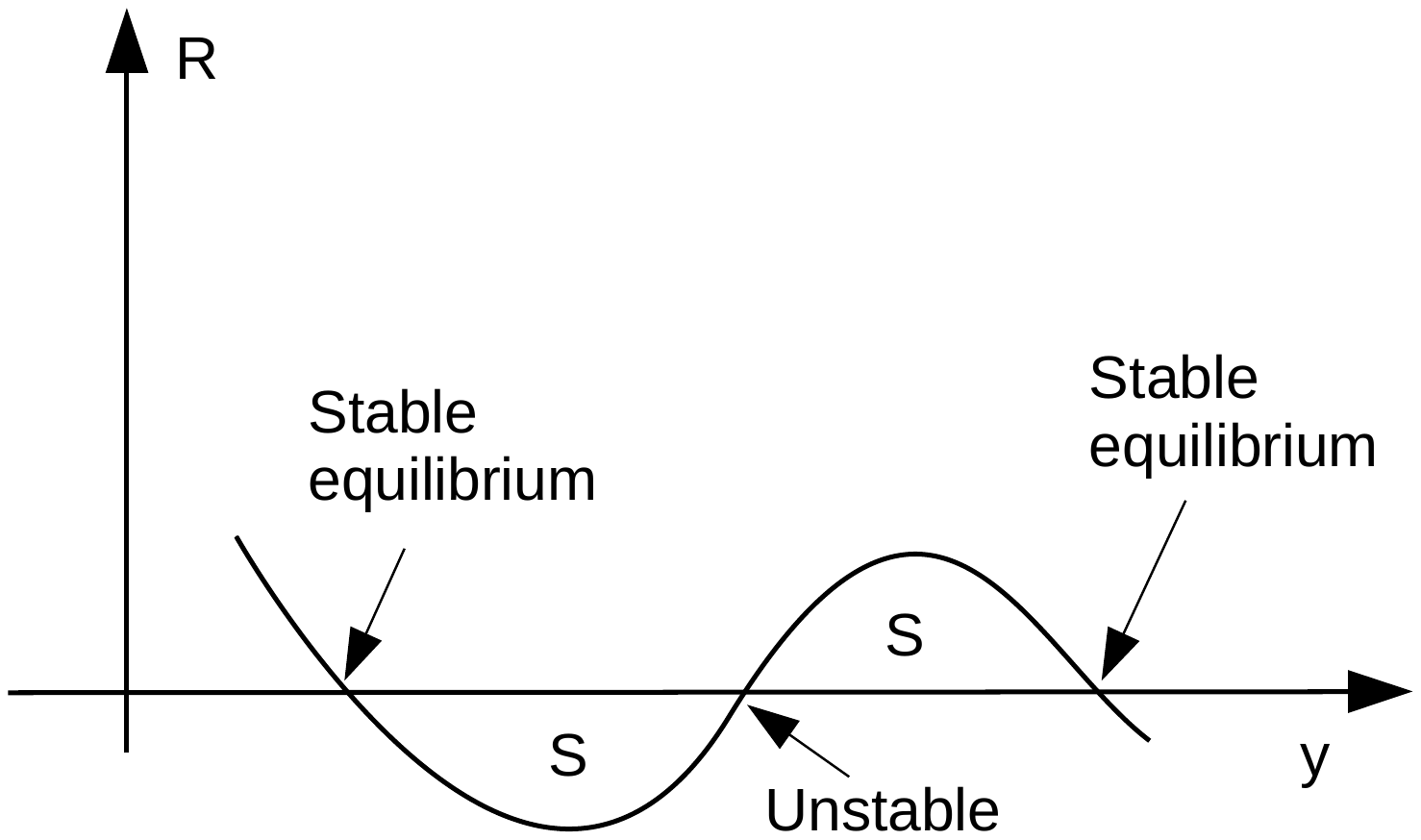}

\caption{Illustration of a Maxwell transition rule, in a form of the production-dissipation
function, $R\left(y\right)$. In order to connect the two stable equilibria,
shown in Fig.\ref{fig:Enstrophy-production-dissipation}, with one
heteroclinic orbit, shown in Fig.\ref{fig:'Oscillator'-pseudo-potential-an},
the areas above and below the abscissa must be equal.\label{fig:Maxwell}}
\end{figure}
For values of parameters $E$ and $b$ other than discussed above,
the solutions given by eq.(\ref{eq:Ysol}) fall in two further categories:
(i) strictly periodic solutions corresponding to $E<E_{*}$, while
$b$ may or may not remain equal to $b_{*}$, Fig.\ref{fig:'Oscillator'-pseudo-potential-an},
and (ii) soliton type solutions when the orbit is homoclinic, starting
and ending at one of the two hyperbolic points. Here $b\neq b_{*}$,
$E=E_{*}$, so we have only one hyperbolic fixed point on the orbit.
Just as the heteroclinic solution described earlier, this solution
also becomes periodic for $E<E_{*}$. The periods of solutions with
$E<E_{*}$ can be calculated using eq.(\ref{eq:Ysol}) as 

\begin{equation}
\mathcal{L}\left(E,b\right)=\frac{\sqrt{2}}{L}\int_{\psi_{1}}^{\psi_{2}}\frac{\partial\varepsilon}{\partial\psi}\frac{d\psi}{\psi\sqrt{\varepsilon\left[E-W\left(\psi\right)\right]}}\label{eq:PeriodOfSC}
\end{equation}
Here the integral runs between the two turning points obtained from
the relation $W\left(\psi\right)=E$. By contrast with compact solutions,
i.e. isolated transitions or solitons (which do not 'fit' into a finite
domain), the periodic solutions can be fully described on a $\left[0,1\right]$
segment, provided that $\mathcal{L}<1$. In general, however, and
especially when $n\mathcal{L}\neq1$ with $n=1,2,\dots$, the boundary
conditions need to be consistent with parameters $b$ and $E$. We
will touch upon this aspect of the solution later.

\subsection{Comparison of time-asymptotic numerical solutions with the analysis\label{subsec:Comparisonasymptotic}}

A comparison of the analytic solutions given above with those obtained
numerically serves several purposes. Firstly, it verifies the code's
accuracy and convergence, while establishing its limitations. The
latter is particularly important in view of the time- and space-scale
disparities inherent in this problem. Secondly, it verifies that the
analytic solutions are indeed the attractors for the time-dependent
solutions. In addition, the numerical integration delineates the basins
of attraction of these analytic solutions. Finally, such comparison
gives insight about a possible evolution of the system beyond the
accessible integration time. 
\begin{figure}
\includegraphics[bb=50bp 0bp 916bp 550bp,clip,scale=0.35]{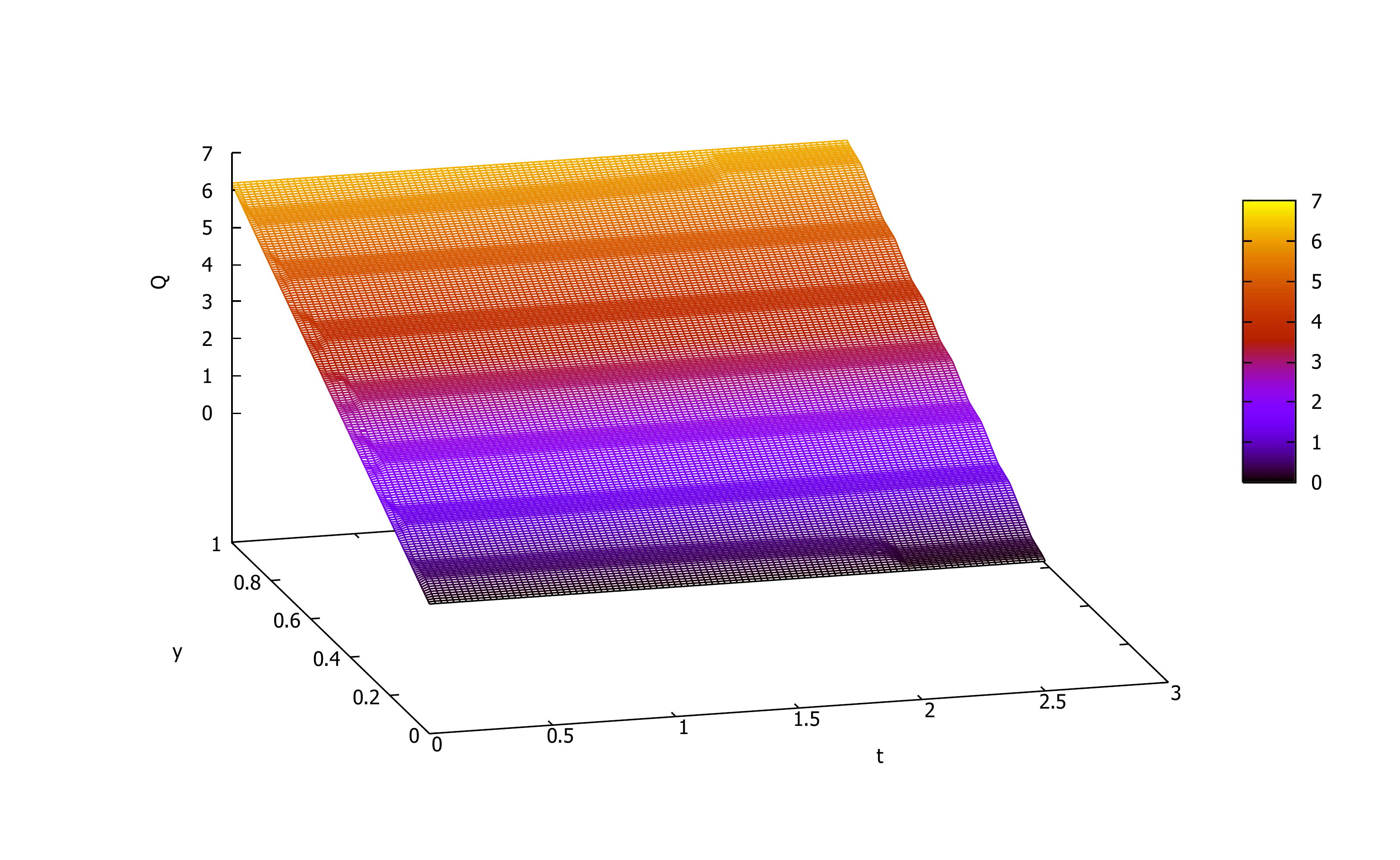}

\caption{Merger of steps shown as a surface plot of the mean vorticity $Q\left(t,x\right)$
with the Dirichlet boundary conditions, $\Delta Q=6.2,$ between the
right and left boundary, $\varepsilon_{0}=10.38$, $L^{2}=1.5\cdot10^{5},$
$D=1.7$. Initially, as many as 12 steps are formed but very rapidly
($t\lesssim0.1)$ they merge to 7. The 7-step configuration lasts
with no numerically significant changes up to the moment $t\approx2$
when two shear layers near the walls disappear and the edge steps
merge into respective walls. The remaining staircase structure persists
up to $t\simeq100$ when the two edge steps merge with their neighbors.\label{fig:Q(x,t)}}
\end{figure}
The latter aspect is crucial in that the time dependent solutions
typically show a long rest - transition burst alternation. Fig.\ref{fig:Q(x,t)}
shows an example of such behavior. After quick ($t\lesssim0.1$) mergers
of the ten out of the twelve initially formed steps ($10\to5$), the
systems sits at the state of seven remaining steps for a long time,
$t\approx2$. Moreover, at this time the staircase merely accommodates
the boundary conditions by attaching each of the two edge steps to
the respective boundary, Fig.\ref{fig:Q(x,t)}. In the next section,
a case of much longer rest will be presented. We will also take a
closer look at a typical merger event. Without analytic predictions,
it is difficult, if not impossible, to tell whether the long rests
are genuine attractors of the system or the next merger is to be expected.
Also, if yes, then what determines the waiting time? Conversely, if
a merger occurs after a long rest period-- during which the profile
agrees with one of the stationary analytic solutions-- one may consider
such a merger as a spurious effect of accumulated round-off errors.
\begin{figure}
\includegraphics[bb=0bp 0bp 612bp 792bp,scale=0.5,angle=270]{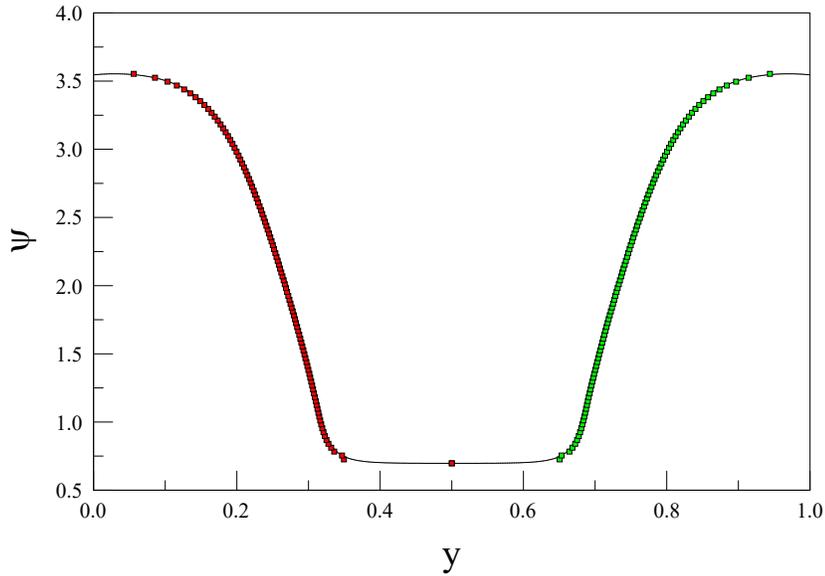}

\caption{Numerical solution of eqs.(\ref{eq:QeqNdim}-\ref{eq:EpsEqNdim})
in long-time asymptotic regime, shown with the solid line. The parameters
are the same as in Fig.\ref{fig:1StepSC}. Exact analytic solution
represented by the two branches of $y\left(\psi\right)$ in eq.(\ref{eq:Ysol}),
shown with red and green squares.\label{fig:FitNumAnal}}
\end{figure}
For the purpose of comparison with numerical solutions, the constants
$b$ and $E$ -- which an analytic solution depends upon -- should
be calculated using boundary conditions imposed on the numerical solutions.
More practically and equivalently, we extract them directly from the
numerical solution already shown in Fig.\ref{fig:1StepSC}. An exact
analytic solution, plotted using eq.(\ref{eq:Ysol}), is compared
to this numerical solution in Fig.\ref{fig:FitNumAnal}. Given the
parameters $b$ and $E$, the analytic solution is obtained by a simple
tabulation of the integral in eq.(\ref{eq:Ysol}), even without regularization
of the integrand singularities at $W=E$. This explains a non-uniform,
coarse sampling, as well as a minor deviation of one point next to
the minimum of $\psi\left(y\right)$ from the numerical curve. The
extrema of $\psi\left(y\right)$ obviously correspond to the algebraic
($E<E_{*}$) or logarithmic ($E=E_{*})$ singularities of $y\left(\psi\right)$
. 

The parameters and boundary conditions for the run in Fig.\ref{fig:FitNumAnal}
are such that the entire profile spans slightly more than one period
of the respective analytic solution, $\mathcal{L}\lesssim1$. Furthermore,
in the asymptotic state shown in the figure, the flux $b$ is constant
throughout the integration domain to within the code accuracy, so
the solution is indeed steady. Note that we varied the code error
tolerance between $10^{-8}-10^{-6}$ (see \citep{MuirBacoli2004}
for the algorithm's description and further references) without sizable
effects on convergence, up to $t\sim100$. Further details on the
comparison of analytical and numerical solutions are given in Appendix
2.

These results -- and particularly the perfect agreement between the
overall analytic and numerical solutions along with the code convergence
(insensitivity to the error tolerance) -- increase our confidence
that the code accurately describes the evolution of the staircase
system in time. 

\section{Staircase merger events\label{sec:Staircase-merger-events}}

The purpose of this section is to connect staircase mergers with the
analytic properties discussed in the previous section. Within the
range of parameters outlined in Sec.\ref{sec:Staircase-Prerequisites},
the numerical integration of eqs.(\ref{eq:QeqNdim}-\ref{eq:EpsEqNdim})
consistently demonstrates the following evolution: long meta-stationary
periods interspersed by quick staircase mergers. During these periods,
the staircase configuration does not change in any noticeable way,
and is accurately described by analytic solutions obtained in the
preceding section. Natural questions then are: what causes the next
merger, and does it become final -- beyond which the staircase configuration
will stay unchanged. Here the word 'unchanged' should be taken with
caution, as our simulations clearly show that nearly perfectly stationary
staircase configurations lasting for $t\gtrsim100$ do merge eventually,
and over times $t\lesssim0.1$. Recall that time here is given in
units of $L^{2}/\sqrt{\gamma}$, eqs.(\ref{eq:EpsEqDim}) and (\ref{eq:Units}).
Despite these remarkably disparate time scales, the answer that we
find below is at least consistent with the analytic solutions.
\begin{figure}
\includegraphics[bb=100bp 0bp 600bp 792bp,scale=0.5,angle=270]{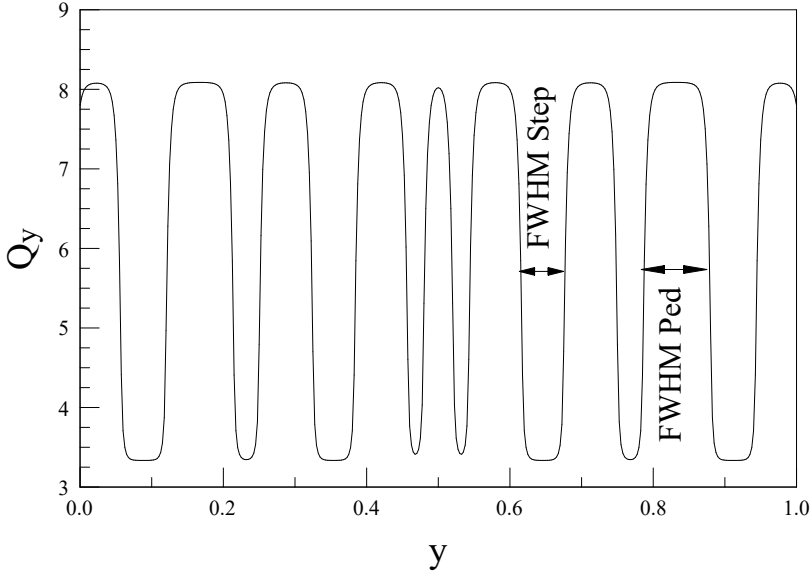}

\caption{Typical profile of $\nabla Q$ (or $Q_{y}$) where FWHM (full width
at half maximum) characterizes the width of shear layers (humps) and
steps (troughs) on the plot. Here $D=1.5,\;\;L^{2}=10^{5},\;\quad\varepsilon_{0}=3.7915,\;\;Q\left(1\right)=6.2,\;\;\varepsilon\left(0,1\right)=10.38$\label{fig:Typical-profile-of_gradQ}}
\end{figure}
Typically, a quasi-stationary staircase forms very quickly ($t\ll1)$
with $n$ steps separated by shear layers exhibiting a significantly
steeper gradient of the mean vorticity $Q_{y}$ also with suppressed
enstrophy level, $\varepsilon$. The number $n$ is determined by
the maximum growth rate (similarly to the results of \citep{Balmforth98},
see also Sec.\ref{sec:Staircase-Prerequisites}) and partly by the
initial conditions. Then, over a somewhat longer time (but still $t<0.1$),
most -- except the boundary-attached -- steps merge with their neighbors.
So, the total number of steps becomes $\approx n/2$. This phase of
the staircase evolution is typically similar to that shown in Fig.\ref{fig:FiveMergers}
(which shows a time history behind the $Q_{y}$ -profile shown in
Fig.\ref{fig:Typical-profile-of_gradQ}). After this initial phase
the staircase persists for a much longer time. It is clear from the
surface plot of the $Q$-flux that it grows rapidly, and deviates
strongly from its globally constant value precisely at the merger
locations. 

\begin{figure}
\includegraphics[clip,scale=0.17]{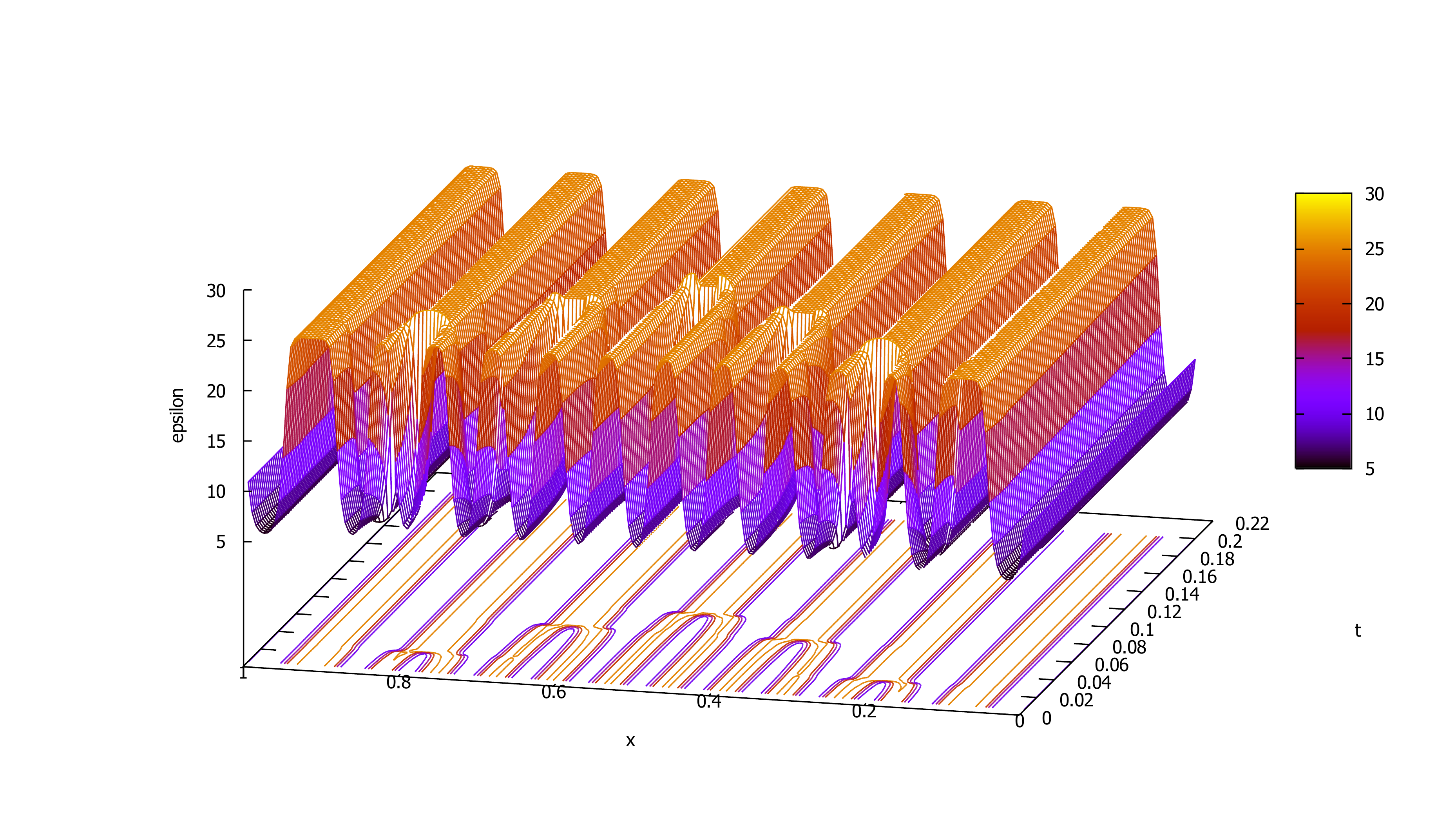}\includegraphics[scale=0.17]{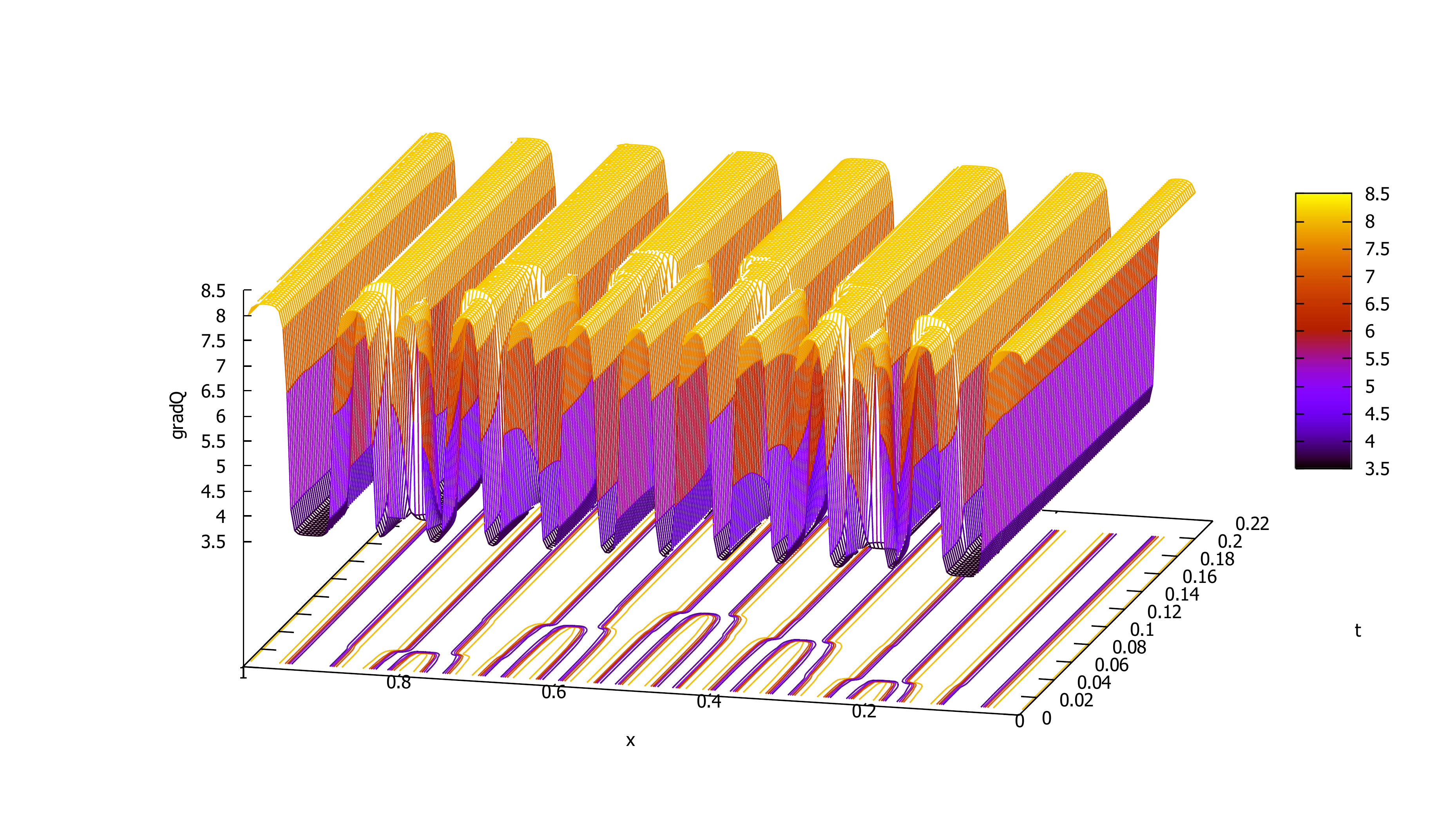}

\includegraphics[scale=0.17]{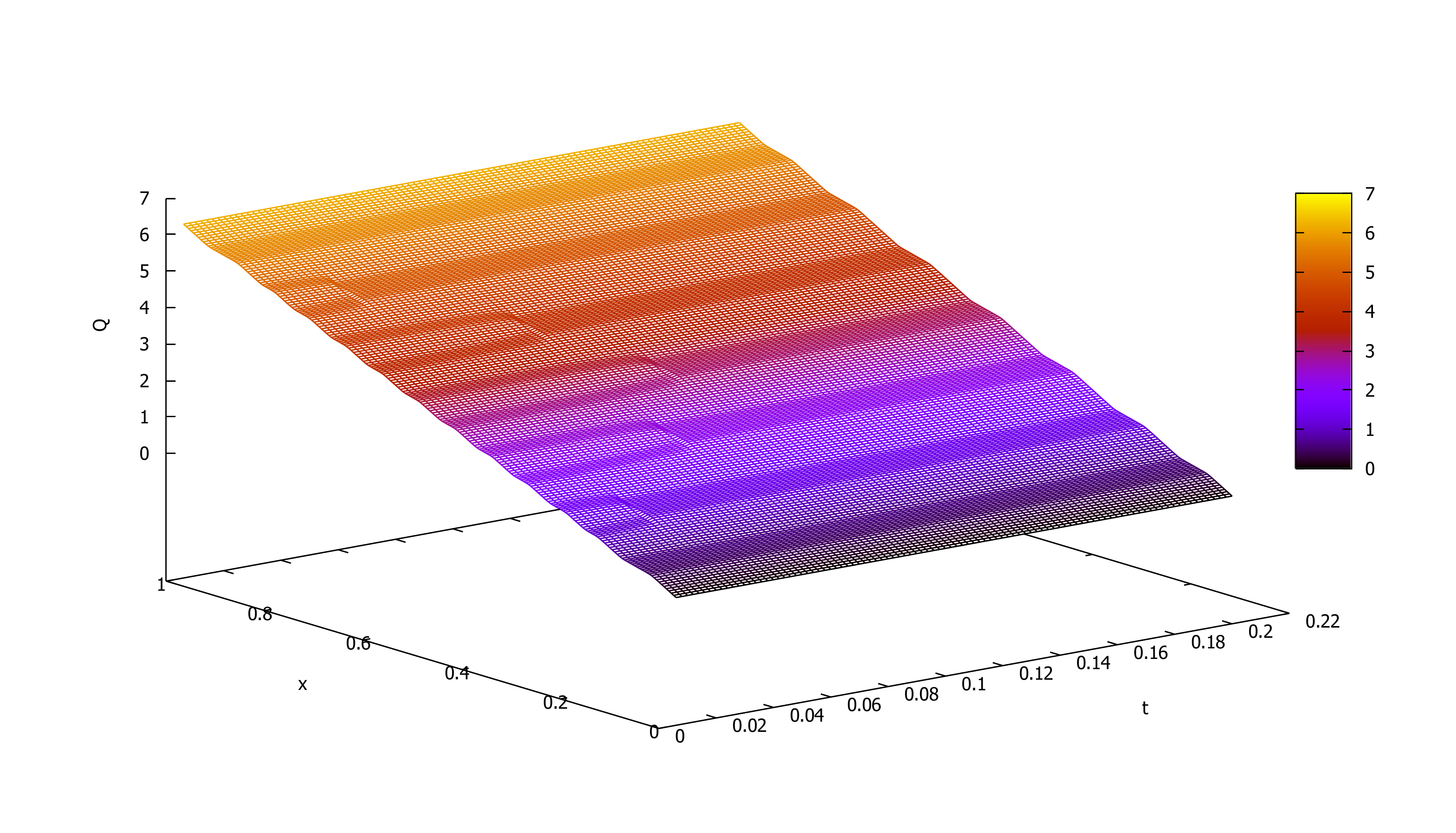}\includegraphics[scale=0.17]{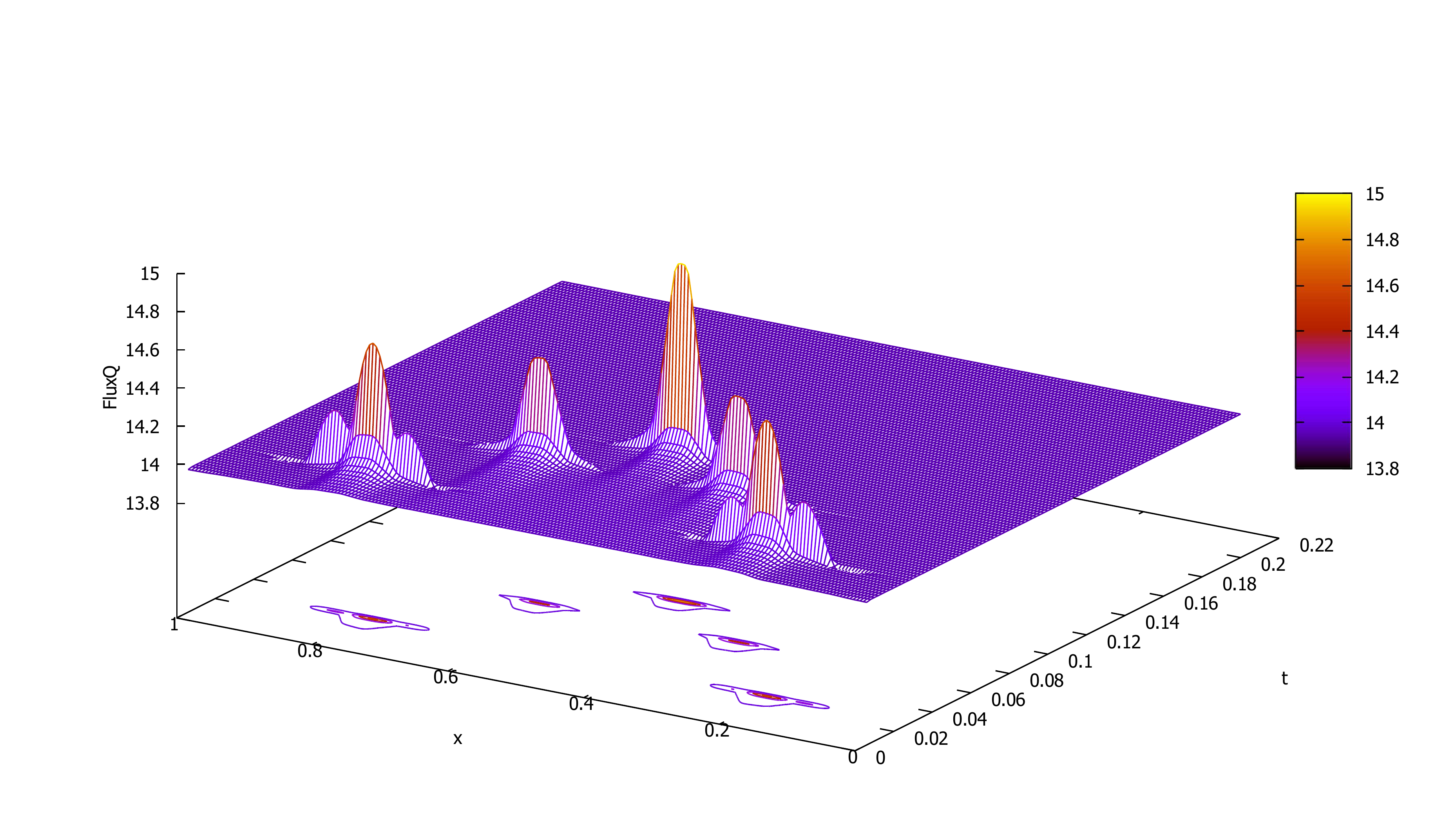}

\caption{Time history and details of the staircase merger sequence that led
to the profile shown in Fig.\ref{fig:Typical-profile-of_gradQ}. The
top two panels show the enstrophy $\varepsilon$ and $Q_{y}$ , while
the bottom panels show the mean potential vorticity $Q$ and its flux
given by eq.(\ref{eq:Qfluxb}). The early relaxation phase $t\lesssim0.01$
is removed from the plots to emphasize the flux strong variation during
the later staircase mergers. \label{fig:FiveMergers}}
\end{figure}
The further fate of a quasi-stationary staircase depends on the following
factors. One factor is the proximity of the staircase to the periodic
solution described in Sec.\ref{sec:Analytic-solution-for}. The portions
of the staircase profile that are periodic do not merge, consistent
with the analytic solutions. As we have seen, all stationary solutions
are exhausted by either periodic or compact (solitary or kink type)
solutions \footnote{The compact solutions technically require an infinite integration
domain, but can be reproduced in a finite domain with exponential
accuracy (see also below).}. Therefore, the periodic segments of the staircase tend to be steady.
An example of such a staircase is shown in Fig.\ref{fig:Q(x,t)}.
After quick initial mergers, this staircase remains periodic in its
interior and only the edge regions deviate from spatial periodicity.
Here, the second factor that influences the evolution of a staircase
enters. This is the boundary effect. Indeed, as we discussed in Sec.\ref{subsec:Comparisonasymptotic},
at $t\approx2$ the edge shear layers disappear and the edge steps
attach themselves to the boundaries. An inspection of the $Q$-flux
shows that it remains constant in the interior where the staircase
is periodic, and it progressively deviates from the constant values
near the boundaries. This ultimately results in the boundary accommodation
event at $t\approx2$, after which the flux returns to its global
average. After this event, however, the staircase becomes non-periodic
at the edges (edge steps are broader than the central ones). As expected,
the edge steps merge with their neighbors, but only at $t\approx100.$ 

A different example of a staircase, that is noticeably \emph{non-periodic,}
is shown in Fig.\ref{fig:Typical-profile-of_gradQ}. Such configurations
tend to merge over significantly shorter time, particularly at those
locations where the steps or shear layers are close to each other.
So, the spacing between the neighboring corners is important for the
mergers. These corners separate steps from shear layers. As we know,
a single corner with a step and a shear layer on each side makes a
stationary structure in an infinite space and is described by a heteroclinic
orbit. It approaches exponentially constant values at $\pm\infty$
that characterize the step and the shear layer, respectively. Two
neighboring corners, for example, form either an isolated step or
isolated shear layer, e.g. Fig.\ref{fig:1StepSC}. Based on Sec.\ref{sec:Analytic-solution-for},
this construction does not belong to any type of stationary solution
in infinite space. However, if its corners are well separated by a
broad step or shear layer, their interaction term is$\sim\exp\left(-L\Delta y\right)\ll1$,
where $\Delta y$ is the distance between the corners and $L\gg1$
(see Sec\ref{sec:Analytic-solution-for} and Fig.\ref{fig:1StepSC}).
So, broad, well separated steps and layers in a staircase must also
persist in time. Typically, when the system relaxes to 4-5, not perfectly
periodic steps, the next merger occurs after several hundreds of time
units. Integration beyond this time would require a significantly
lower error tolerance. Under these circumstances, analytic predictions
become much more reliable than numerical ones.

\begin{figure}
\includegraphics[clip,scale=0.5,angle=270]{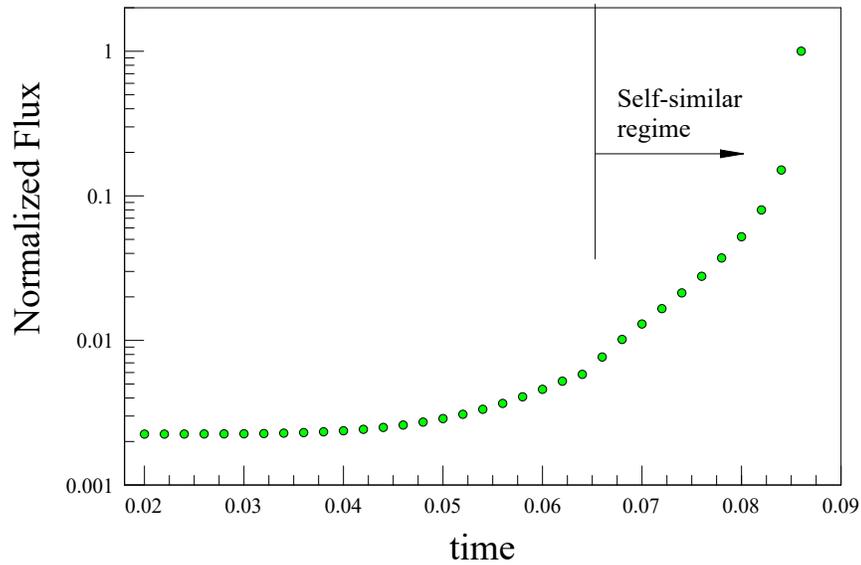}

\caption{$Q-$ flux as a function of time immediately before and during the
merger event near $t\approx0.09$ at the middle of integration domain
at $y=0.5$ shown in Fig.\ref{fig:FiveMergers}. For the clarity of
representation and a functional fit, shown in Fig.\ref{fig:Fit-of-the-Flux},
the flux is adjusted by subtracting a reference value $F_{B}\approx13.96088$
which is close to a globally averaged flux.\label{fig:FluxMergeMid}}
\end{figure}
\begin{figure}
\includegraphics[clip,scale=0.5,angle=270]{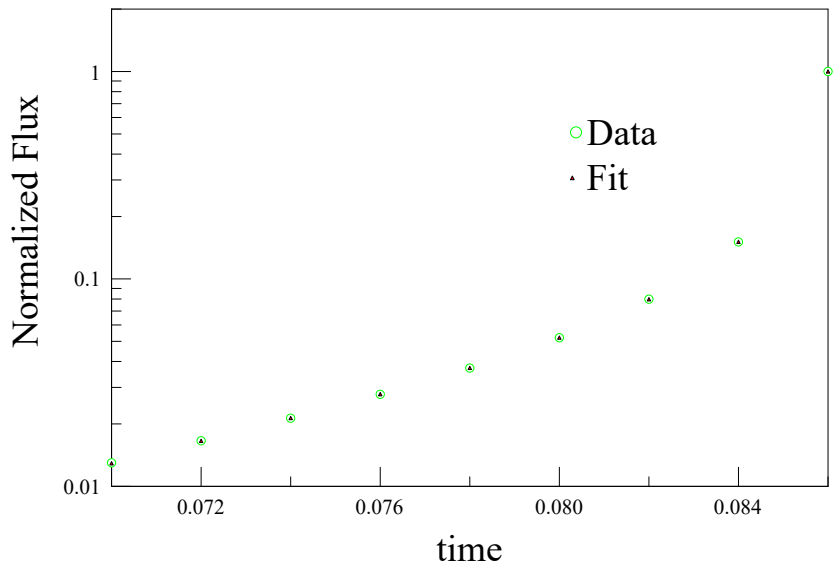}

\caption{Fit of the flux in its self-similar phase of evolution for $0.070<t<0.086$.
The fit is given by $F=F_{0}+B/\left(t_{0}-t\right)^{\alpha}$ with
$t_{0}\approx0.0863$, $B\approx0.000806$, $\alpha\approx0.879$,
subtracted background $F_{B}\approx13.96088$, a residual flux constant
$F_{0}\approx-0.0171$.\label{fig:Fit-of-the-Flux}}
\end{figure}

Staircase steps merge in a rather interesting way. Fig.\ref{fig:FiveMergers}
shows a sequence of such mergers of 12 initial steps (see the $Q$-panel).
The mergers proceed symmetrically from the boundaries towards the
center of the integration domain. This process continues until the
mergers converge at the center and the central two steps merge into
a bigger step. Here, we concentrate on this last merger by zooming
into it in time. The best variable for characterization of the merger
is the $Q-$ flux, which is shown in Fig.\ref{fig:FluxMergeMid} at
the point $y=0.5$ as a function of time. The flux remains constant
when no mergers occur (constant $b$ in stationary solutions of Sec.\ref{sec:Analytic-solution-for}),
so we subtract this constant value for clarity and for the purpose
of functional fitting below. As seen from the plot, the flux builds
up in two distinct phases before it drops abruptly to its averaged
value after the merger. The first phase is an initial growth that
lasts to about $t\approx0.065$. The flux increase remains relatively
small, $\lesssim0.01$. The second phase is clearly explosive and
can be accurately fit by the following function, Fig.\ref{fig:Fit-of-the-Flux}

\[
F=F_{0}+B/\left(t_{0}-t\right)^{\alpha}
\]
with $t_{0}\approx0.0863$, $B\approx0.000806$, $\alpha\approx0.879$,
and a residual flux $F_{0}\approx-0.0171$ . Note that apart from
this last constant, the background value $F_{B}\approx13.9609$ has
been subtracted from the total flux. By contrast with the initial
phase, the local flux excess grows to a value $\gtrsim1$ before the
merger occurs, and the total flux then drops abruptly to its background
value. 

\section{conclusion\label{sec:Discussion-and-conclusions}}

The Phillips's staircase \citep{Phillips72} has long since outgrown
its original context. Indeed, it is of general interest as an outcome
of the nonlinear scale and pattern selection processes in systems
with an homogeneously mixed conserved density. The present paper is
devoted to numerical studies and analysis of a limiting case of new
staircase model recently introduced in \citep{Ashourvan2017}, in
the context of geostrophic fluids and magnetized plasmas. It evolves
the potential vorticity (PV) and potential enstrophy with a versatile
model of mixing set by the Rhine-scale. The model controls enstrophy
production by the PV gradient, additional independent pumping, and
nonlinear damping. These system properties render it bistable, which
is the key to staircase formation. 

The main results of the current investigation of this model can be
summarized as follows:
\begin{enumerate}
\item parameter regimes and mixing length scalings in which the staircases
occur are analytically constrained to satisfy: $l_{0}^{2}/l^{2}=\left(1+l_{0}^{2}/l_{R}^{2}\right)^{\kappa}$,
$\kappa>1$, $l_{R}$ is the Rhines scale $l_{R}^{2}=\varepsilon/Q_{y}^{2}$
\item staircase formation, their properties and stability were studied numerically,
for differing parameters and boundary conditions
\item analytic solutions for steady state staircase configurations are obtained
and categorized into periodic and isolated (compact) type solutions
\item the time of persistence of a quasistationary staircase configuration
is demonstrated to depend entirely on its proximity to the closest
stationary solution
\item staircase mergers are studied and the likelihood of the next merger
event is elucidated, again, from the perspective of the configuration's
proximity to the closest periodic solution
\item merger dynamics is identified as explosive in time, and localized
in space to a pair of staircase neighboring elements (steps or jumps)
\item step mergers produce localized bursts of PV mixing and transport
\item analytic solutions provide insight into the meta-stationary staircase
structures beyond the capability of numerical integration
\end{enumerate}
While meta-stationary staircase configurations, which persist practically
unchanged between the merger events, can be fully understood using
analytic solutions, merger dynamics requires further quantitative
studies. In this paper, we give only a phenomenological account of
staircase merging. We note that recovering mergers and understanding
merger dynamics are somewhat problematic. Mergers are observed in
numerical solutions of the basic hydrodynamic equations \citep{srinivasan2012zonostrophic}
and in models \citep{Ashourvan2016PhRvE,Ashourvan2017} but not in
gyrokinetic simulations \citep{Dif-Prad2010}.

This paper has elucidated in depth the basic physics of the ``Hasegawa-Mima''
\citep{HasegawaM1978,SagdeevConvC1978} staircases, in which fluctuations
are produced by external stirring, and which makes no distinction
between density and potential (i.e. $\tilde{n}/n_{0}=\left|e\right|\hat{\phi}/T_{e}$
for $m\neq0$ fluctuations). A more interesting case is the ``Hasegawa-Wakatani''
\citep{HasegawaW1987} staircase, for which:
\begin{enumerate}
\item density and potential evolve separately, but are coupled.
\item drift wave instability processes allow access to free energy and promote
the growth of vorticity flux at the expense of particle flux.
\item consistent with ii.), vorticity and particle fluxes are dynamically
coupled
\item spontaneous transport barrier formation is possible. 
\end{enumerate}
Staircase formation in the Hasegawa-Wakatani system are discussed
in Refs.\citep{Ashourvan2017,HahmDiamondTurbTransp2018}. These studies
are primarily computational. Further analysis will be coming in a
future publication. 

\section*{Acknowledgements}

We thank G. Dif-Pradalier, Y. Hayashi, D.W. Hughes and W.R. Young
for helpful discussions. P.H. Diamond acknowledges useful conversations
with participants in the 2015 and 2017 Festival de Theorie, the 2018
Chengdu Theory Festival and the 2014 Wave-Flows Program at KITP. This
research was supported by the Department of Energy under Award No.
DE-FG02-04ER54738.

\appendix

\section{Staircase Parameters\label{sec:Appendix1}}

An extremum condition for eq.(\ref{eq:ExtremaFofKsi}), $F^{\prime}\left(\xi\right)=0$,
contains only one parameter, $\varepsilon_{0}$, and can be written
as

\[
\left(\xi+1\right)^{3}-2\varepsilon_{0}\xi=0
\]
It is convenient to write the solution of this cubic equation in a
trigonometric form. The two positive roots $\xi_{0,1}$ are given
by

\begin{equation}
\xi_{n}=2\sqrt{\frac{2\varepsilon_{0}}{3}}\sin\left[\frac{1}{3}\sin^{-1}\left(\frac{3}{2}\sqrt{\frac{3}{2\varepsilon_{0}}}\right)+\frac{2}{3}\pi n\right]-1,\;\;\;n=0,1\label{eq:KsiN}
\end{equation}
From this expression, we obtain a condition for the existence of the
two isolated roots $\xi_{0,1}$, namely $\varepsilon_{0}>27/8$ (the
third roots is irrelevant here). Another requirement for the bistability
is imposed on $Q_{y}$ by the two relations in eq.(\ref{eq:QyBistInt1}).
However, this constraint is not precise in that $Q_{y}$ also changes
in the transition from one stable state to the other, along with $\xi$.
At the same time, one can easily determine $Q_{y}\left(\xi\right)$
variation by assuming that it changes in space but remains stationary.
This assumption relates $Q_{y}$ to $\xi$ by the constant diffusive
flux in eq.(\ref{eq:QeqNdim}). It is totally justified for quasi-stationary
phases of the SC evolution, Sec.\ref{sec:Analytic-solution-for}.

\section{Comparison between analytical and numerical solutions\label{sec:Apendix2}}

Periodic analytic solutions with small conventional diffusivity and
a moderate number of total steps are characterized by the values $E$
and $b$ that are very close to their separatrix values $E_{*}$ and
$b_{*}$, eqs.(\ref{eq:FirstInt}) and (\ref{eq:WofPso}). This results
in extended flat parts of the $\psi\left(y\right)$ profile in Fig.\ref{fig:FitNumAnal}
and those of $\varepsilon$ and $Q_{y}$, shown earlier in Fig. \ref{fig:1StepSC}.
That $E\approx E_{*}$, is also obvious from the expression for the
staircase period in eq.(\ref{eq:PeriodOfSC}). On writing it as $\mathcal{L}=I\left(E\right)/L$
and noting that $L\gg1$ and $\mathcal{L}\sim1$, we find that $I\gg1$
which implies $E\approx E_{*}$ (recall that $\max W\left(\psi\right)=E_{*}$).
Moreover, since $I\sim\ln\left(E_{*}-E\right)^{-1}$ and $L\approx48$
in this particular run, $E$ may, in principle, become close to $E_{*}$
to the machine accuracy, because $E_{*}-E\sim\exp\left(-L\right)$.
By examining the numbers we find that the difference between the numerical
and analytical values of $\min\psi\left(y\right)$, shown in Fig.\ref{fig:FitNumAnal}
is $\sim10^{-4}$. This result implies that $E_{*}-E\sim10^{-8}$
which is not inconsistent with the above estimate of the logarithm
of this quantity.

\bibliographystyle{C:/Users/mmalkov/bibtex/prsty}
\bibliography{PaperPRF1}

\end{document}